\documentclass[final,english]{elsarticle}
\usepackage[T1]{fontenc}
\usepackage[latin9]{inputenc}
\usepackage{varioref}
\usepackage{units}
\usepackage{amsmath}
\usepackage{amssymb}
\usepackage{graphicx}
\usepackage{esint}
\usepackage{caption}
\usepackage{subcaption}

\pdfpageheight\paperheight
\pdfpagewidth\paperwidth

\journal{Elsevier}
\usepackage{upgreek}
\usepackage[bookmarks,bookmarksopen,bookmarksdepth=6]{hyperref}
\usepackage{cases}
\usepackage{url}
\usepackage{lineno}

\usepackage{a4wide}

\usepackage{newtxtext,newtxmath,amsmath}
\makeatother

\usepackage{babel}
\begin{document}

\title{Determination of the electric field in highly-irradiated silicon sensors using edge-TCT measurements}
\author[]{R.~Klanner$^1$ \corref{cor1}}
\author[]{G.~Kramberger$^2$}
\author[]{I.~Mandi\' c$^2$}
\author[]{M.~Miku\v z$^2$}
\author[]{M.~Milovanovi\' c$^3$}
\author[]{J.~Schwandt$^1$}

\cortext[cor1]{Corresponding author. Email address: Robert.Klanner@desy.de.}

\address{$^1$Institute for Experimental Physics, University of Hamburg, Luruper Chaussee 149, D\,22761 Hamburg, Germany}
\address{$^2$ Jozef Stefan Institute, Jamova 39, Sl\,1000 Ljubljana, Slovenia}
\address{$^3$ Jozef Stefan Institute, now at DESY Zeuthen, Platanenallee 6, D\,15738 Zeuthen, Germany}


\begin{abstract}

 A method is presented which allows to obtain the position-dependent electric field and charge density by fits to velocity profiles from edge-TCT data from silicon strip-detectors.
 The validity and the limitations of the method are investigated by simulations of non-irradiated $n^+p$~pad sensors and by the analysis of edge-TCT data from non-irradiated $n^+p$~strip-detectors.
 The method is then used to determine the position dependent electric field and charge density in $n^+p$ strip detectors irradiated by reactor neutrons to fluences between 1 and $10 \times 10^{15}$~cm$^{-2}$ for forward-bias voltages between 25~V and up to 550~V and for reverse-bias voltages between 50~V and 800~V.
 In all cases the velocity profiles are well described.
 The electric fields and charge densities determined provide quantitative insights into the effects of radiation damage for silicon sensors by reactor neutrons.

\end{abstract}


\maketitle
 \tableofcontents
 \pagenumbering{arabic}

 \section{Introduction}
  \label{sect:Introduction}
 The knowledge of the electric field and of the effective doping distribution in silicon sensors is essential to predict and understand their performance.
 Although qualitatively the physics of the change of the electric field with hadron irradiation was understood already about 25 years ago~\cite{Eremin:1995,Eremin:2002}, a reliable quantitative determination is still lacking.
 The attempts to calculate the electric field in irradiated sensors using TCAD simulations with trap parameters from spectroscopic measurements have been only partially successful so far~\cite{Lutz:1999, Petasecca:2006, Dalal:2014, Moscatelli:2016, Schwandt:2018}.
 Difficulties are the large number of radiation-induced states~\cite{Pintilie:2009} with frequently only poorly known properties, the problem of implementing cluster defects in the simulation~\cite{Donegani:2018} and the influence of high electric fields on the defect properties.
 Thus, a method to determine experimentally the electric field is highly desirable.

 In Ref.~\cite{Kramberger:2014} $n^+p$~strip sensors irradiated with hadrons up to fluences of $10^{16}$~cm$^{-2}$ were investigated using the edge-Transient Current Technique (edge-TCT).
 In edge-TCT~\cite{Kramberger:2010} the sensor is illuminated by a focussed near-infrared laser beam, which is  parallel to the sensor surface and normal to the strips, and the current transients are recorded as a function of the distance of the laser beam from the readout plane.
 From the initial part of the induced current pulse, which is proportional to the sum of the hole and electron velocities, the electric field at the position of the laser beam can be determined.
 In Ref.~\cite{Kramberger:2014} a parametrisation of the electric field (Fig.~5 of~\cite{Kramberger:2014}) has been assumed to extract the parameters of the model from the data.
 As a cross-check the electric field thus determined has been used to simulate the position dependence of the initial currents.
 Although an approximate description has been achieved, significant differences are observed.
 In the present paper a method is proposed, which allows determining the electric field by directly fitting the initial currents without assumptions on the shape of the electric field.
 The method is applied to the data of Ref.~\cite{Kramberger:2014} as well as to reference data for the non-irradiated strip-sensor.

 The following section discusses the sensors, their irradiation and the edge-TCT measurements.
 Then, the method used to extract the electric field from the data is derived, the assumptions made discussed, and the fitting procedure presented.
 To demonstrate the validity of the model, as a first step data from a non-irradiated sensor are analysed and the results compared to the expectations for a non-irradiated pad detector.
 Then the model is used to determine the position dependence of the electric field in the sensor as a function of forward and reverse bias voltage and irradiation fluence.
 Finally the results are discussed and suggestions for further studies made.

  \section{Sensors, irradiation and measurements}
  \label{sect:Sensors}

 The sensors investigated are $p$-type AC-coupled micro-strip detectors fabricated on float-zone silicon by Hamamatsu Photonics~\cite{HPK}.
 The main properties of the sensors are listed in Table~\ref{tab:Sensor}.
 More information is given in Refs.~\cite{Kramberger:2014, Kramberger:2010, Unno:2011}.
 The isolation between the $n^+$-implants is achieved by a $p$-spray implant with $2 \times 10^{12}$~B-ions/cm$^2$, and a single $p$-stop implant with $8 \times 10^{12}$~B-ions/cm$^2$.
 This information is relevant for estimating the electric field in-between the strips close to the Si-SiO$_2$ interface.

 \begin{table} [!ht]
  \centering
   \begin{tabular}{c|c|c|c|c|c|c|c}
     thickness & crystal& pitch    & $n^+$ width & strip length & Area & B-doping  & $U_{fd}$ \\
     $ \upmu$m &orientation & $\upmu$m & $\upmu$m         & cm   & cm$^2$      & cm$^{-3}$ & V \\
     \hline
      300       &$\langle 100 \rangle$ & 100      & 20    & 0.8  &0.62 & $2.8 \times 10^{12}$ & 180 \\

   \end{tabular}
    \caption{Sensor parameters.
     For the B-doping and the full depletion voltage, $U_{fd}$, the values obtained from the fits for the non-irradiated strip detector (Sect.~\ref{sect:non-irradiated}) are given.
    \label{tab:Sensor}}
  \end{table}

  One strip of the sensor was connected via a current amplifier (10~kHz--1~GHz) to a 1.5~GHz sampling scope.
  The remaining strips were connected via 50~$\Omega $ resistors to a fixed potential (virtual ground).
  For each measurement the average of 400 transients was recorded.

  The sensor was illuminated through a carefully polished edge with the light from a laser with a wavelength of 1064~nm and a pulse width of 50~ps.
  At this wavelength the light absorption length in non-irradiated silicon is about 1~mm at room temperature~\cite{Green:2008}.
  It decreases with increasing temperature and with increasing fluence, when irradiated by hadrons,~\cite{Scharf:2019}.
  The laser light was focussed to a full-width-half-maximum of about $7~\upmu$m at the position of the readout strip, and its position  was controlled with sub-micron precision by a computer-controlled moving stage.

  For the measurements analysed in this paper, the sensors were irradiated with neutrons in the TRIGA reactor of JSI in Ljubljana to 1 MeV neutron-equivalent fluences, $\Phi _{eq} $, of (1, 2, 5, 10)$~\times 10^{15}$~cm$^{-2}$~\cite{Snoj:2012, Zontar:1999}.
  The fluence uncertainty is estimated to be below 10~\%.
  After irradiation, the sensors were annealed for 80~min at $60^{~\circ }$C.
  In addition, data from a non-irradiated sensor were analysed.
  For the irradiated sensors the measurements were taken at $ - 20^{~\circ }$C, and for the non-irradiated one at $ + 20^{~\circ }$C .

  \begin{figure}[!ht]
   \centering
    \includegraphics[width=0.5\textwidth]{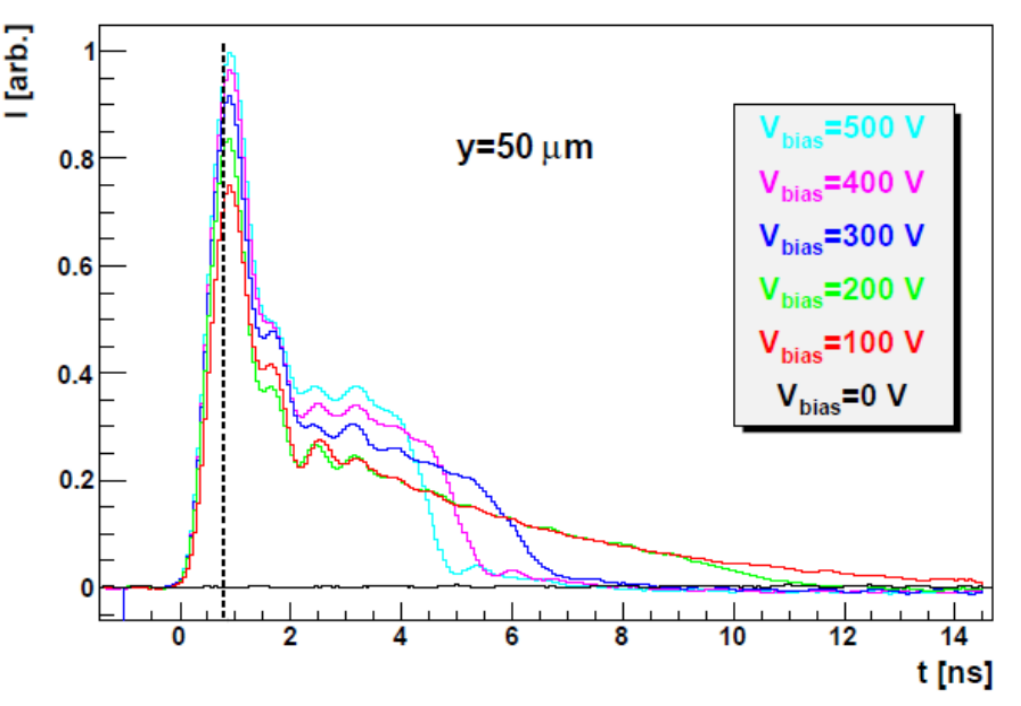}
   \caption{Measured current transients from the non-irradiated sensor at $20^{~\circ }$C with the laser beam $50~\upmu$m from the strips (from \cite{Kramberger:2014}).}
  \label{fig:Trans_nonirr}
 \end{figure}

 Fig.\,\ref{fig:Trans_nonirr} shows the measured current transients for the non-irradiated sensor with the laser positioned $50~\upmu$m from the readout strip for reverse bias voltages between 0 and 500~V.
 The initial peak is dominated by the current induced by electrons, which move in the high field towards the readout electrode.
 The tail at later times is caused by the holes, which move in the decreasing field towards the $p^+$~back side of the sensor.
 The rise time of the transient reflects the electronics response function of the system sensor and readout.

 As discussed in more detail in Sect.~\ref{sect:AppendixB}, there are different ways of obtaining \emph{velocity profile}s, $vel$, from the data.
 Two examples are:
 The integration of the transient up to the time $t_{int}$, and the maximum of the time derivative of the initial pulse.
 As shown in Ref.~\cite{Kramberger:2014}, up to $t _{int} = 800$~ps the shape of $vel$ is independent of $t _{int}$.
 Shorter values of $t _{int}$ decrease the sensitivity of $vel$ to charge trapping but increase the relative fluctuations.
 For the analysis of this paper $vel$ is obtained from the maximal slopes of the current transients.
 For the sensors irradiated to the fluences $\Phi _{eq}$ the velocity profiles $vel(y; U, \Phi _{eq})$, have been measured as a function of the distance $y$ from the strip plane in $5~\upmu$m steps and of the applied voltage $U$ up to 800~V for the irradiated sensors, and up to 500~V for the non-irradiated sensor.
 In addition, the dark current $I_{dark}(U, \Phi _{eq})$ has been recorded.

  \section{Model and fit procedure}
  \label{sect:Model}

\begin{figure}[!ht]
   \centering
    \includegraphics[width=0.4\textwidth]{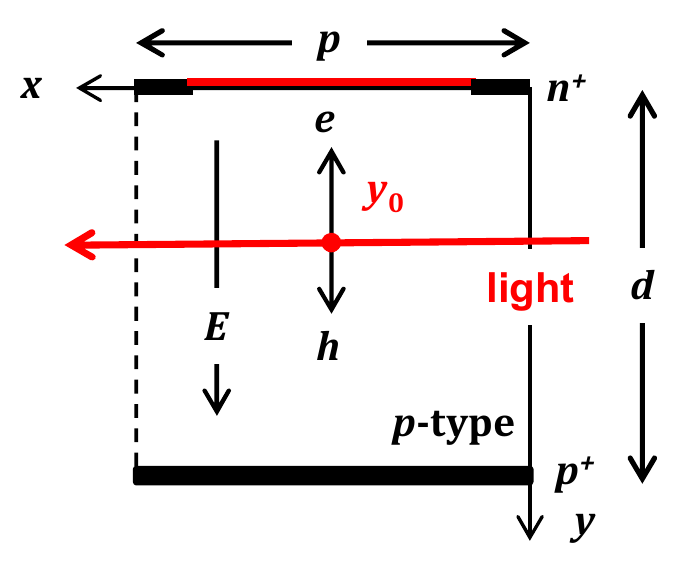}
   \caption{Schematic drawing of the cross section of the strip sensor and the coordinate system used.
    The region between the centers of two readout strips is shown.
    The strip pitch is $p$, the sensor thickness $d$, and the length of the strips in the $z$-direction is 8~mm.}
  \label{fig:Sensor}
 \end{figure}

 In this section the method used to extract the electric field from $vel(y)$ is presented and the  assumptions discussed.
 Fig.~\ref{fig:Sensor} shows the sketch of the cross section of the sensor and defines the coordinate system.

  A charge $Q$ moving with the velocity $\vec{v}\big(\vec{r}(t)\big)$ along the trajectory $\vec{r}(t)$ induces in the readout electrode the current
   \begin{equation}\label{equ:Igeom}
    I(t) = Q \cdot \vec{v}\big(\vec{r}(t)\big) \cdot \overrightarrow{E_{w}}\big(\vec{r}(t)\big),
   \end{equation}
  where $\overrightarrow{E_{w}}(\vec{r})$ denotes the weighting field~\cite{Riegler:2018}, which has the units 1/cm.
  In general, a time-dependent weighting vector is required to describe $I(t)$.
  However, as demonstrated  in Ref.~\cite{Schwandt:2019}, time-independent weighting fields can be used to describe the signal of fully depleted non-irradiated sensors and for sensors irradiated by hadrons to $\Phi _{eq} \gtrsim 10^{13}$~cm$^{-2}$.
  The weighting field is obtained as the difference of the electric field of the biased sensor plus $\Delta U = 1$~V on the readout electrode minus the electric field of the biased sensor divided by $\Delta U$.
  For a light beam traveling at constant $y$ in the $x$-direction, the current is induced by all charges generated along the light path.
  If the light attenuation length is long compared to the strip pitch and the strip length long compared to the sensor thickness, $d$,  Eq.~\ref{equ:Igeom} can be used to describe the induced current.
  In this case $\overrightarrow{E_{w}}(\vec{r}) = \hat{e}_y /d $, with the unit vector in the y-direction $\hat{e}_y$, and
  $Q$  the moving charge per unit length multiplied with the strip pitch, $p$.

  This can be seen in the following way:
  If the dielectric constant in the sensor is uniform and independent of the electric field, the effective weighting field for the charge distribution $\mathrm{d}^3Q / \mathrm{d}x\mathrm{d}y \mathrm{d}z = q_0 \cdot \delta(y-y_0) \cdot \delta(z) / p$ is $\overrightarrow {\mathcal{E} _w}(\vec{r}) = 1/p \cdot \int _{- \infty } ^{+ \infty} \overrightarrow{E}_{w}(\vec{r})~\mathrm{d}x $.
  From symmetry arguments follows that for a sensor with an infinite extension in the $x$- and $z$-directions the integral does not depend on $x$ and $z$.
  Symmetry arguments also require that the $x$- and $z$-components of $\mathcal{E}_w$ cancel, and that only the $y$-component is finite.
  Changing the voltage on the readout electrode by 1 V does not change the charge density inside the sensor, and from Gauss' law follows that $\overrightarrow {\mathcal{E} _w}$ is independent of $y$.
  From $ \int _0 ^d \mathcal{E}_w(y)~\mathrm{d}y = 1$  follows the result
  $\overrightarrow {\mathcal{E} _w}(\vec{r}) = \hat{e}_y/d$.

  The change of $\overrightarrow {\mathcal{E}_w}$ due to the finite attenuation length of the near-infrared light used for the measurements is discussed in Sect.~\ref{sect:AppendixA}.

 The drift velocities of the charge carriers are $\vec{v}_i(\overrightarrow{E}, T) = \mu _i (E,T) \cdot \overrightarrow{E}$, where the parametrisation of the hole mobility, $\mu _h$, and of the electron mobility $\mu _e$ for electrons for $\langle 100 \rangle$ silicon from Ref.~\cite{Scharf:2015} are used; $T$ is the temperature.
 Although there is some evidence that the low-field mobilities decrease after hadron irradiations for fluences $\Phi _{eq} \gtrsim  10^{15}$~cm$^{-2}$~\cite{Scharf:2018, Masetti:1983}, fluence-independent $\mu _i$~values are assumed.

  \begin{figure}[!ht]
   \centering
    \includegraphics[width=1.0\textwidth]{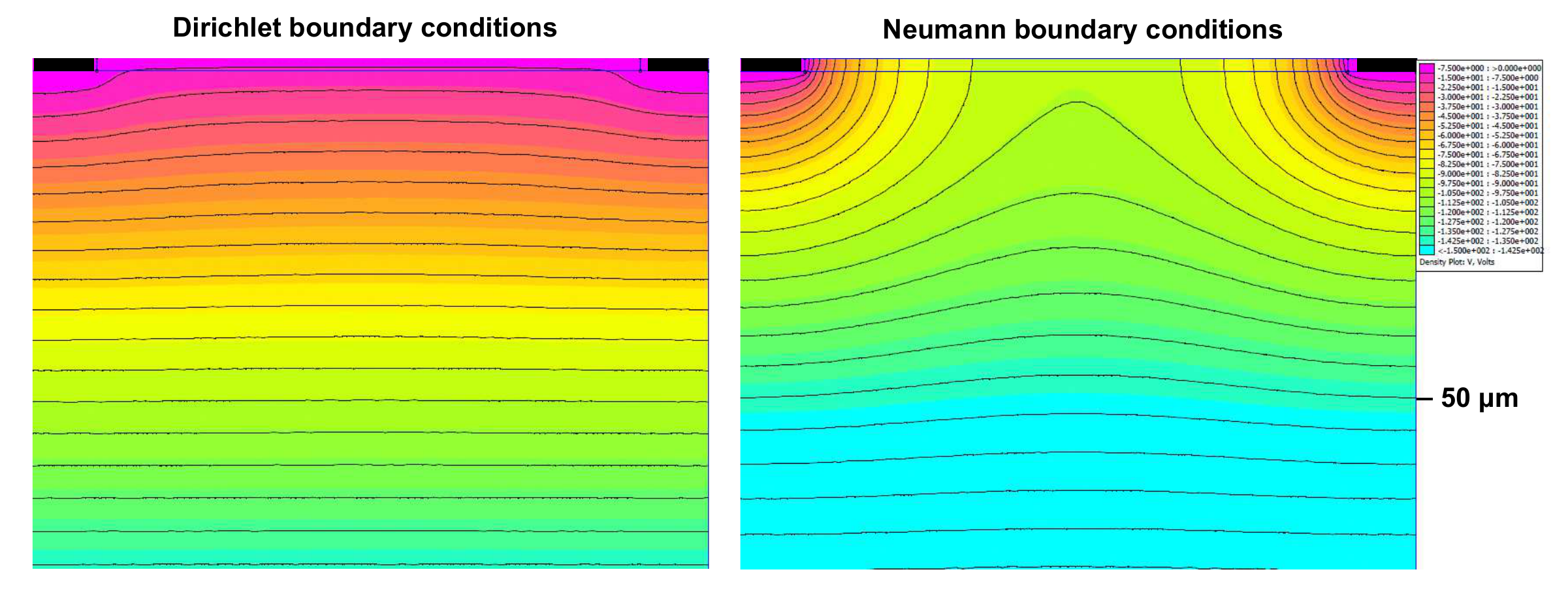}
   \caption{Simulated potential, $\Phi $, in a strip detector with the parameters given in Table~\ref{tab:Sensor} close to the readout strips biased to $-400$~V.
    Between the strips there is a $2~\upmu$m thick SiO$_2$ layer.
    The results for Dirichlet boundary conditions (same $\Phi$ on top of the SiO$_2$ as on the strips) and for Neumann boundary conditions ($\partial \Phi /\partial y = 0$ on top of the SiO$_2$) are shown. (Colour on-line)}
  \label{fig:Potential}
 \end{figure}

 For the electric field  $\overrightarrow{E}\big(\vec{r}\big) = E(y) \cdot \hat{e}_y $ is assumed.
 The electric field close to the strip plane critically depends on many parameters, some of which are only poorly known.
 It depends on the  dark current, the $p$-spray and $p$-stop doping, the charge densities at the Si-SiO$_2$~interface and the electrical boundary condition on the outer SiO$_2$ surface.
 Simulations using the software of Ref.~\cite{Meeker:2018} for two extreme conditions, Dirichlet and Neumann boundary conditions, no dark current and no Si-SiO$_2$~interface charges are shown in Fig.~\ref{fig:Potential}.
 For the Dirichlet boundary condition (potential on the outer Si-SiO$_2$ equal to the strip potential), the electric field is similar to the field of a pad diode, and the assumption that it has only a y-component which is independent of $x$  is a good approximation.
 However, it is a very poor description for the Neumann boundary condition (zero of the derivative of the potential normal to the boundary).
 In Refs.~\cite{Poehlsen:2013, Poehlsen1:2013, Schwandt:2017} the electric field for a $p^+n$~strip detector in the region close to the plane of the readout strips  has been studied as a function of ionising  dose, biasing conditions and time after switching on the voltage.
 It has been found, that the boundary condition changes with time because of the finite surface resistivity of the SiO$_2$, which is a strong function of humidity and temperature.
 Depending on the humidity, time constants varying  between one hour and several days have been measured at room temperature.
 In Ref.~\cite{Schwandt:2017} simulations of the changes of the electric field as a function of the time after biasing a segmented $n^+p$-sensor are given.
 In the data analysis presented in this paper, the effect of the lack of knowledge of the field close to the strip plane is investigated by performing fits starting at different minimal $y$-values.

 Next the fit procedure is described, which is used to extract the electric field, $E(y)$, from the measured velocity profile.
 For the fit $n_E$ electric field values, $E_i$, at the equidistant positions $y_i$ are assumed, which are free parameters of the fit.
 For $E$ at other $y$~positions a linear inter- and extrapolation is used, and the constraint $\int _0 ^d E(y)~\mathrm{d}y = U$ is applied.
 The bias voltage is denoted $U$.
 The values of the velocity profile measured at the positions $y_k$ are denoted $vel_k$.
 The corresponding values of the model are $u_k = \big( \mu_h(E_k) + \mu_e(E_k) \big) \cdot E_k$.
 For the fit a $y$-range with $n_k$ values of $vel_k$ is selected, and
   \begin{equation}\label{equ:Chisq}
    \chi ^2 = \frac{1} {\sigma _{vel}^2} \sum _{k=1} ^{n_k} \bigg( 1-\frac{vel_k} {vscale \cdot u_k} \bigg)^2
     + w_{pen} \sum _{i = 2} ^{n_E -1} \bigg( \frac{0.5 \cdot (E_{i-1} + E_{i+1}) - E_i} {E_i} \bigg) ^2
   \end{equation}
 as a function of the parameters $E_i$ and \emph{vscale} is minimised.
 The scale factor $vscale$ is introduced to normalise the $u_k$~values to the $vel_k$~values.
 The relative uncertainty of the $vel_k$~values is denoted $\sigma_{vel}$, for which 2~\% is assumed, and $w_{pen}$ is a penalty term, which prevents excessive fluctuations of adjacent $E_i$~values.
 The choice of the value of $w_{pen}$ is not too critical.
 A value was chosen, which increases the $\chi ^2$ of the fit by $\approx 50$~\% compared to the value for $w_{pen} = 0$.


  \section{Electric field of the non-irradiated sensor for reverse bias}
  \label{sect:non-irradiated}

\begin{figure}[!ht]
   \centering
    \includegraphics[width=0.8\textwidth]{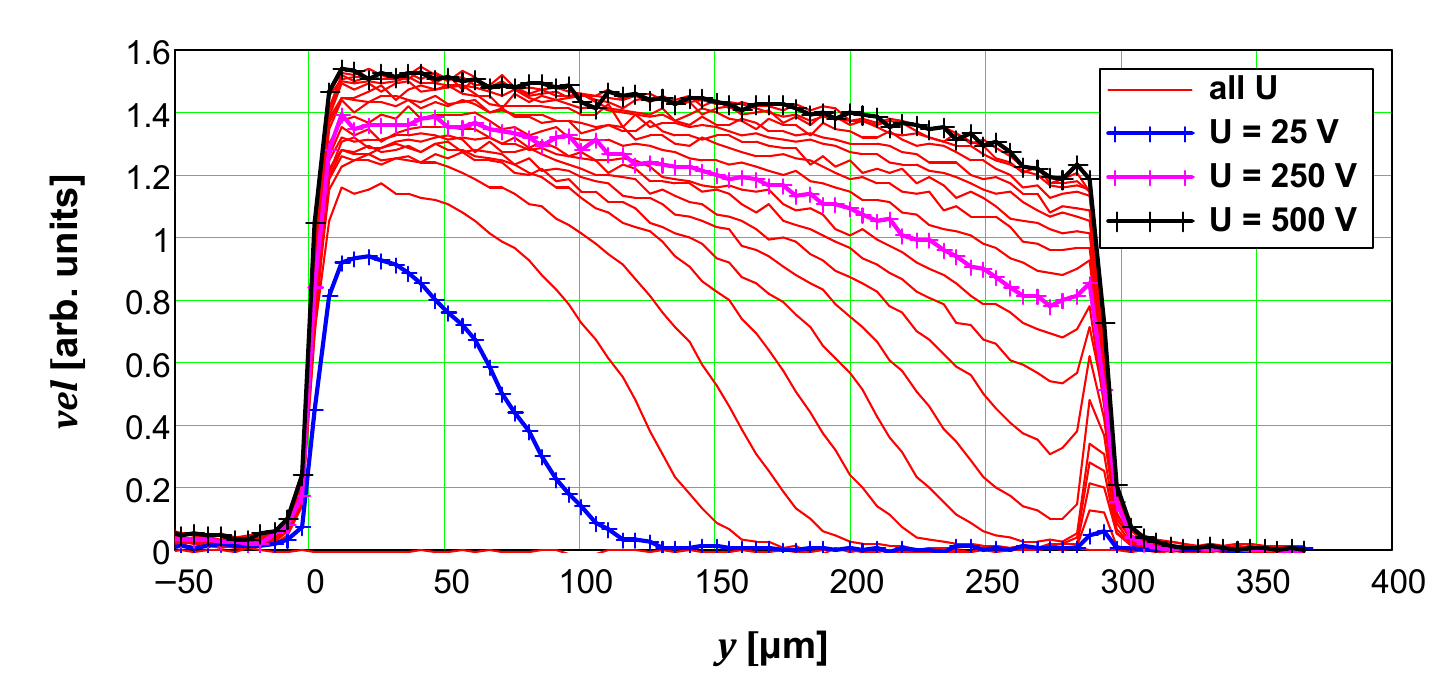}
   \caption{Velocity profiles, $vel$, from Ref.~\cite{Kramberger:2014} of the non-irradiated strip sensor measured at $+20~^\circ $C for reverse voltages between 25 and 500~V in 25~V steps.
   For a given position $y$, $vel$ increases with increasing bias voltage $U$.}
  \label{fig:Vel-non}
 \end{figure}

 To test the analysis method, as a first step the velocity data of the non-irradiated strip-sensor from Ref.~\cite{Kramberger:2014} are analysed.
 The sensor parameters are given in Table~\ref{tab:Sensor}.
 The measurements were taken at $+20~^\circ$C for 25 reverse voltages between 25 and 500~V.
 The measured velocity profiles are shown in Fig.~\ref{fig:Vel-non}.

 The velocity profiles, $vel$, directly reflect the electric field and the field dependence of the hole and electron mobilities.
 The maximum field is at the $n^+p$~junction, which is close to $y = 0$.
 This is also the region where $vel$ has its maximum.
 Below full depletion, signals are only generated by the charges produced in the depletion region, which expands to higher and higher $y$~values with increasing bias voltage $U$.
 At low voltages the slope in the falling part of $vel$, $\mathrm{d}vel/\mathrm{d}y$, is approximately independent of $U$.
  This is expected as the slope of the electric field, $\mathrm{d}E/\mathrm{d}y$, is constant for a uniform doping and at low fields the carrier velocities are approximately proportional to the field.
 For $U \gtrsim 180$~V the full depletion of the sensor is reached, and charges are collected from the entire sensor.
 With increasing $U$ the velocity profiles flatten, which reflects the decrease of the mobilities  with electric field.
 In the region of the $p^+$ backside implant close to $y = d \approx 300~\upmu$m, a spike in $vel$ is observed, in particular for values of $U$ in the vicinity of the full-depletion voltage.
 It is caused by the electric field from the diffusion of holes from the $p^+$~region into the $p$~region.
 The effect decreases at low voltages, because of the finite resistivity of the non-depleted silicon, which results in a time dependent weighting field as discussed in Ref.~\cite{Riegler:2018}.
 The narrow width of the spike as well as the sharpness of the decrease of $vel$ at the two sides of the sensor, directly reflects the width of $7~\upmu$m FWHM of the laser beam focus.

 \begin{figure}[!ht]
   \centering
   \begin{subfigure}[a]{0.5\textwidth}
    \includegraphics[width=\textwidth]{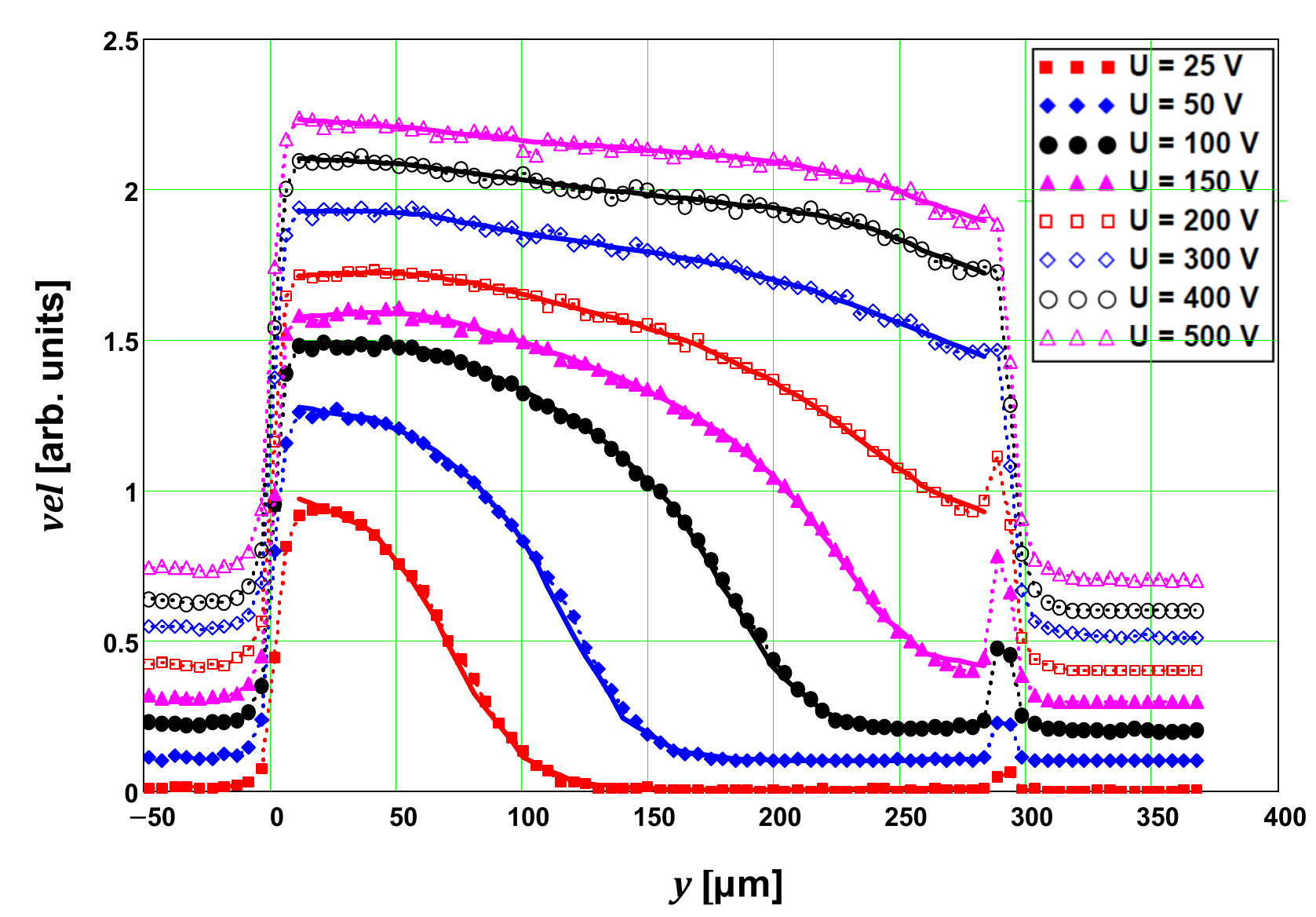}
    \caption{ }
    \label{fig:Velfit-non}
   \end{subfigure}%
   \begin{subfigure}[a]{0.5\textwidth}
    \includegraphics[width=\textwidth]{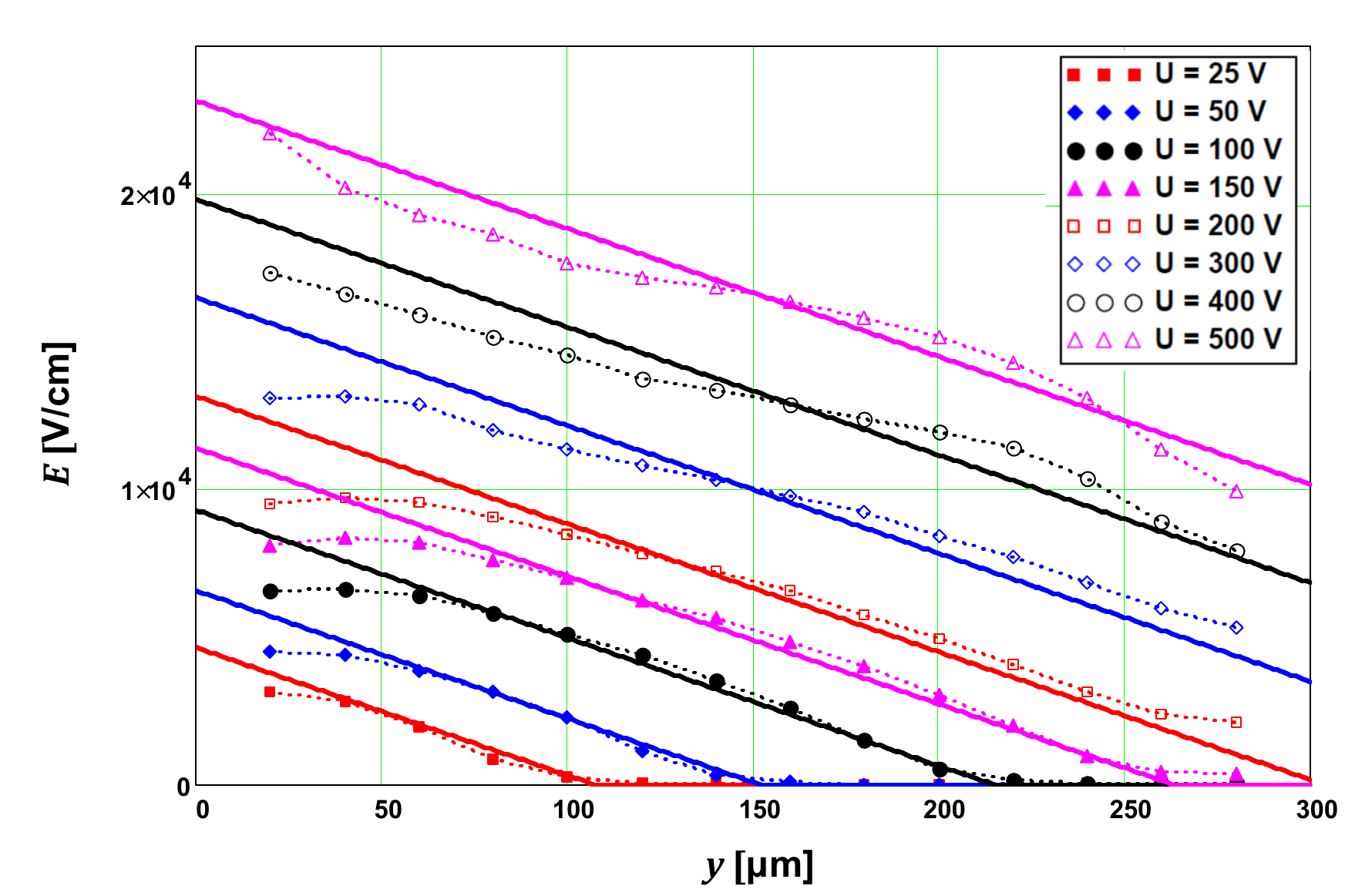}
   \caption{ }
    \label{fig:Efit-non}
   \end{subfigure}%
   ~
   \caption{
     (a) Results of the fits to the velocity profiles for the non-irradiated sensor for different voltages.
         For clarity, the individual data are shifted vertically by steps of 0.1.
         The symbols are the experimental data $vel_k$ at the position $y_k$ and the lines the results of the model fit using Eq.~\ref{equ:Chisq}.
     (b) Electrical field values $E_i$ at the positions $y_i$ from the fits (symbols).
      The dotted lines indicate the linear interpolation used to calculate the velocity profiles of the model shown as lines in (a).
      The solid lines are the electric fields expected for a pad diode with uniform doping.
    }
  \label{fig:Fit-non}
 \end{figure}

  Next the fits of the $vel$~data by the model discussed in Sect.~\ref{sect:Model} are presented.
 The number of $E$~values to be fitted is chosen to be $n_E = 14$, and the $y$~positions of the $E_i$~values are $y_i = i \cdot 20~\upmu$m, with $i = 1,~2,..., ~n_E$ i.~e. between $20~\upmu$m and $280~\upmu$m.
 Fig.~\ref{fig:Fit-non} shows the results of the fits to the data in the interval $y = 11.6$ to $283.5~\upmu$m.
 It is seen that the model provides a good description of the measurements, with a value of the $\chi ^2$ as expected from the statistical fluctuations of the data.
 Fig.~\ref{fig:Efit-non} shows for selected $U$~values the results for $E(y)$ (symbols) and the corresponding electric fields of a pad diode with similar doping (solid straight lines).
 Typical differences are below $\pm 10$\%.
 As expected, the biggest differences are  observed for $ y \lesssim 50~\upmu$m, where the influence of the strips, the surface boundary conditions and the electronics response function are important.
 Actually, the differences are smaller than expected from the discussions of Sections.~\ref{sect:Model} and \ref{sect:AppendixB}.

 At low voltages, where the sensor is only partially depleted, the electric field determined by the fit reproduces the linear $y$~dependence and the expected depletion depth.
 At high voltages and small $y$~values the electric fields are high and the drift velocities approach saturation.
 As a result the accuracy of the field determination degrades and larger differences are observed.
 Nevertheless, the results are satisfactory and demonstrate the validity of the method.

 To study the effects of the non-uniform fields at small $y$~values (see Fig.~\ref{fig:Potential}) and of the electronics response function, fits for different $y$~ranges are performed.
 The results obtained when varying the lower $y$~value between $11.6~\upmu$m and $51.3~\upmu$m, and the upper $y$~value between $273.5~\upmu$m and $283.5~\upmu$m, give equally good descriptions of the data and compatible $E$~values.

  \begin{figure}[!ht]
   \centering
    \includegraphics[width=0.65\textwidth]{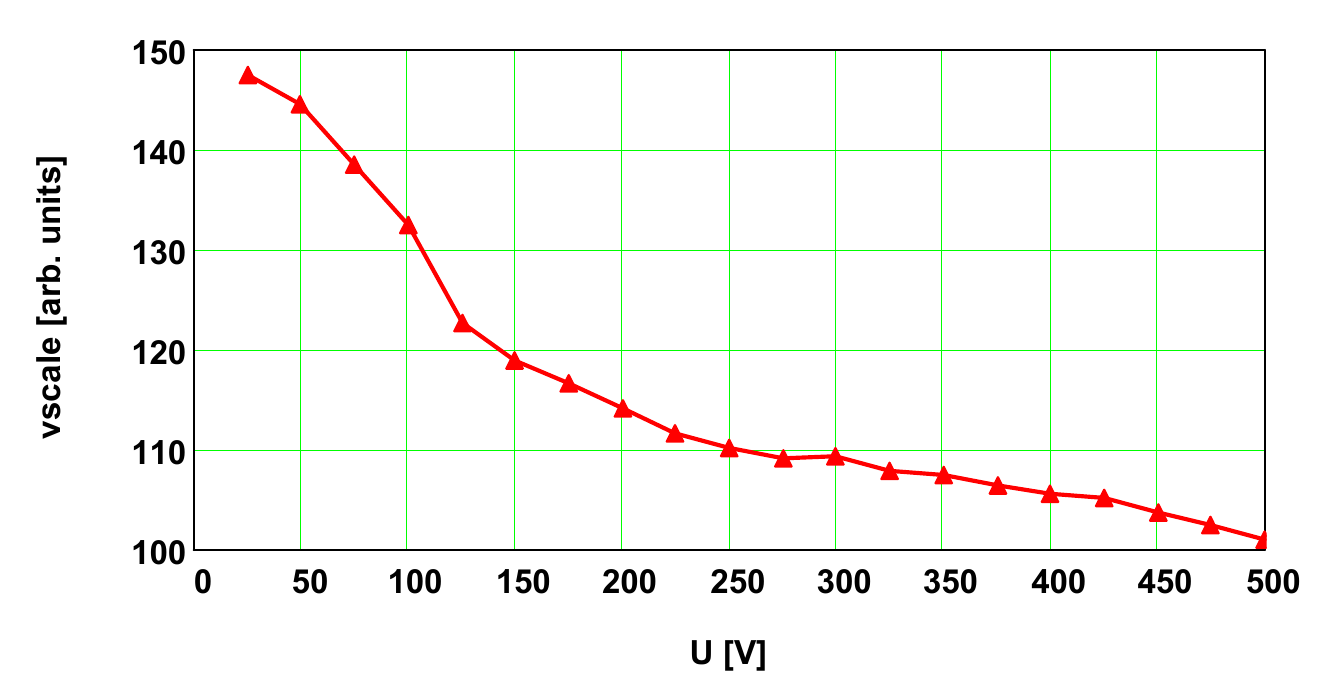}
    \caption{The scaling factor $vscale (U)$ for the non-irradiated detector.}
   \label{fig:Vscale-non}
  \end{figure}

 Fig.~\ref{fig:Vscale-non} shows the voltage dependence of $vscale(U)$, with which the signal from the model calculation has to be multiplied in order to describe the velocity profiles.
 It has to be introduced, as the relation between the initial slope of the pulse, which is used to determine the velocity profiles, and the actual velocities of the charge carriers, is not known with sufficient accuracy.
 An increase of \emph{vscale} with decreasing $U$ is expected from the finite rise time, $t_r$, of the initial current transient.
 As discussed in Sect.~\ref{sect:AppendixB}, a finite $t_r$  causes a decrease of $vel$ compared to the value for $t_r = 0$ close to the boundary of the sensor.
 For the measurements of this paper, the effects become significant for $y \lesssim 40~\upmu$m and for $y \gtrsim 270~\upmu$m.
 The absolute decrease is approximately independent of $U$,  however, the relative decrease is biggest at low $U$.
 As a result \emph{vscale} has to increase  with decreasing $U$ in order to satisfy the condition $U = \int _0 ^d E~dy$.

 To summarise this section:
 The model proposed in Sect.~\ref{sect:Model} provides an excellent description of the velocity data from the edge-TCT measurements of the non-irradiated silicon strip detectors in the 25~V to 500~V range of the measurements.
 The extracted electric field distributions agree approximately with the expectation for non-irradiated silicon strip detectors.
 In particular the voltage dependence of the depletion depth and the linear decrease of the electric field with the distance from the $n^+p$-junction are well described.
 The scale factor between the model and the velocity, $vscale$, depends on the voltage, which is understood only qualitatively.
 The results obtained for the non-irradiated silicon strip detectors are sufficiently encouraging, so that the method is used in the next  section to extract the electric fields in highly-irradiated silicon strip detectors.

  \section{Electric field of the irradiated sensors}
   \label{sect:Eirr}

  \subsection{Forward bias}
   \label{sect:irr-forw}

  In this section the  analysis of the $vel$ data Ref.~\cite{Kramberger:2014} for forward voltages and three neutron fluences are presented.
  The voltage ranges are $25 - 300$~V for $\Phi _{eq} = 2 \times 10^{15}$~cm$^{-2}$, $25 - 500$~V for $\Phi _{eq} = 5 \times 10^{15}$~cm$^{-2}$ and $25 - 400$~V for $\Phi _{eq} = 10^{16}$~cm$^{-2}$.
  The data were recorded at $- 20 ~ ^\circ$C.

\begin{figure}[!ht]
   \centering
   \begin{subfigure}[a]{0.5\textwidth}
    \includegraphics[width=\textwidth]{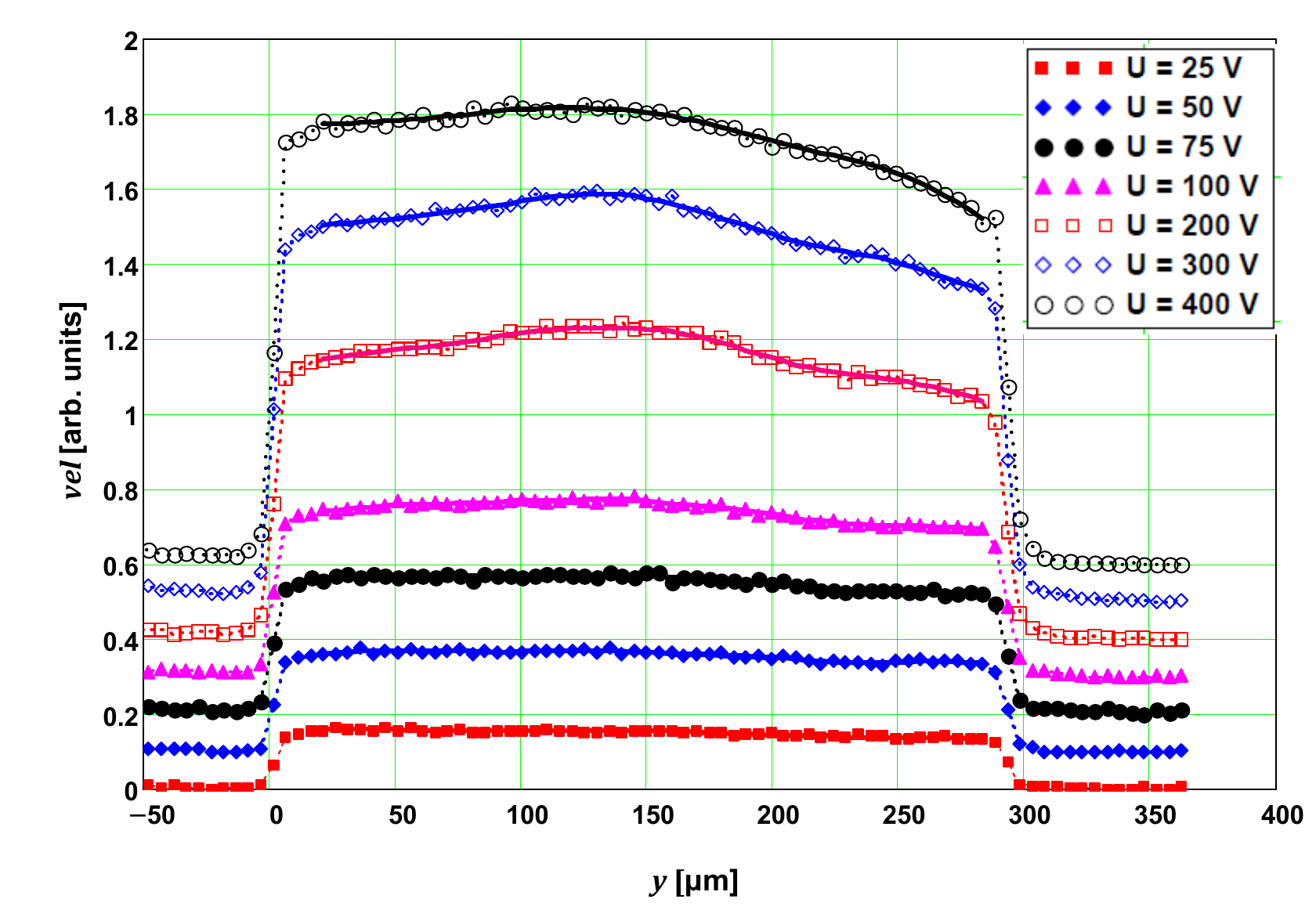}
    \caption{ }
    \label{fig:Vel-irr-forw-5E15}
   \end{subfigure}%
   \begin{subfigure}[a]{0.5\textwidth}
    \includegraphics[width=\textwidth]{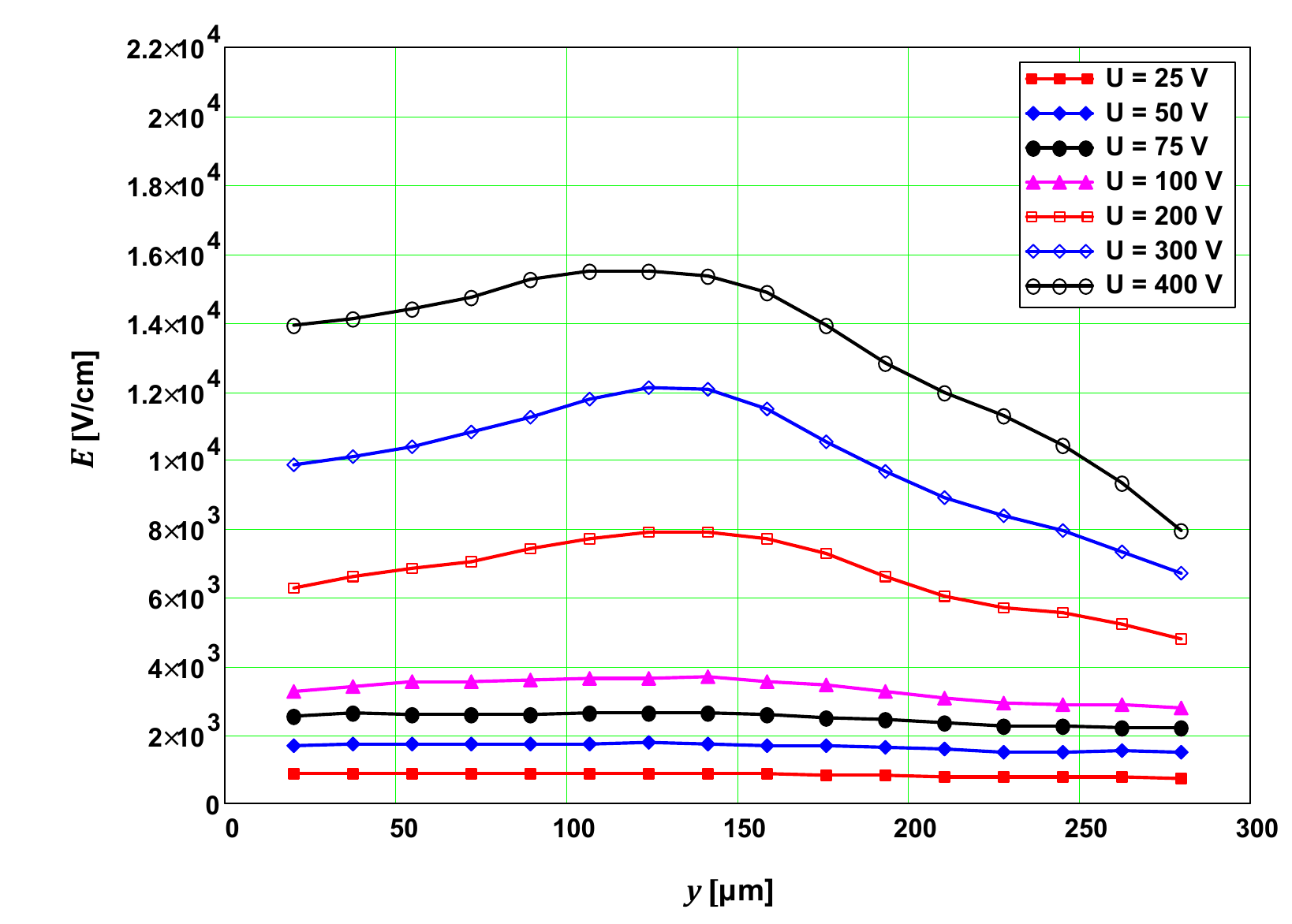}
   \caption{ }
    \label{fig:Efit-irr-forw-1E16}
   \end{subfigure}
   \caption{
     (a) Fits to the velocity profiles for the sensor irradiated to $\Phi _{eq} = 10^{16}$~cm$^{-2}$  for forward bias.
         For clarity, the individual data are shifted vertically by steps of 0.1.
         The symbols are the experimental data and the solid lines, which are hardly visible because of the good description of the data, are the results of the model fit.
     (b) Electrical fields  from the fits to the velocity profiles.
         The lines indicate the linear interpolation used to calculate the velocity profiles of the model shown as lines in (a).
    }
  \label{fig:Fit-1E16-forw}
 \end{figure}

\begin{figure}[!ht]
   \centering
   \begin{subfigure}[a]{0.5\textwidth}
    \includegraphics[width=\textwidth]{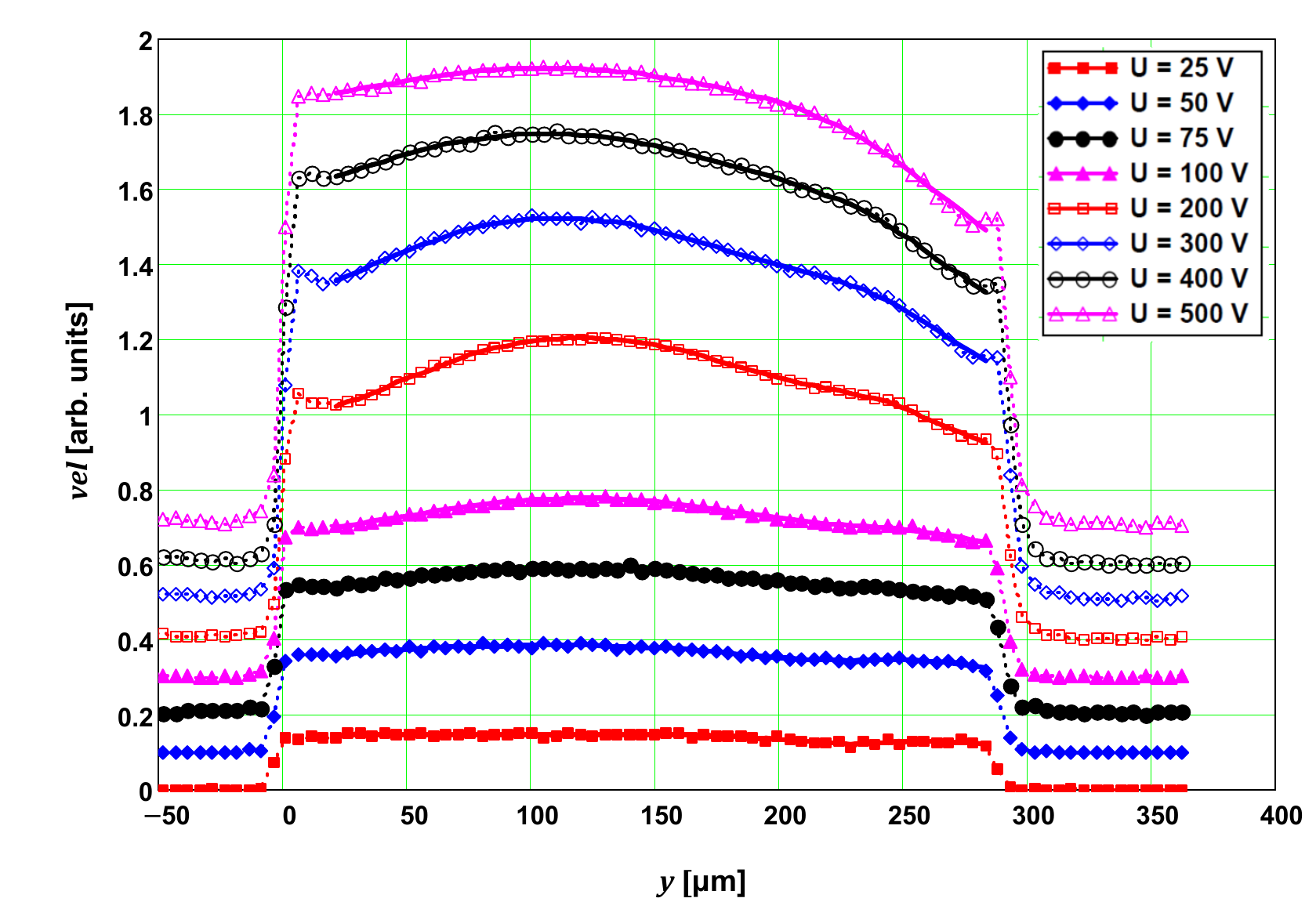}
    \caption{ }
    \label{fig:Vel-irr-forw-5E15}
   \end{subfigure}%
   \begin{subfigure}[a]{0.5\textwidth}
    \includegraphics[width=\textwidth]{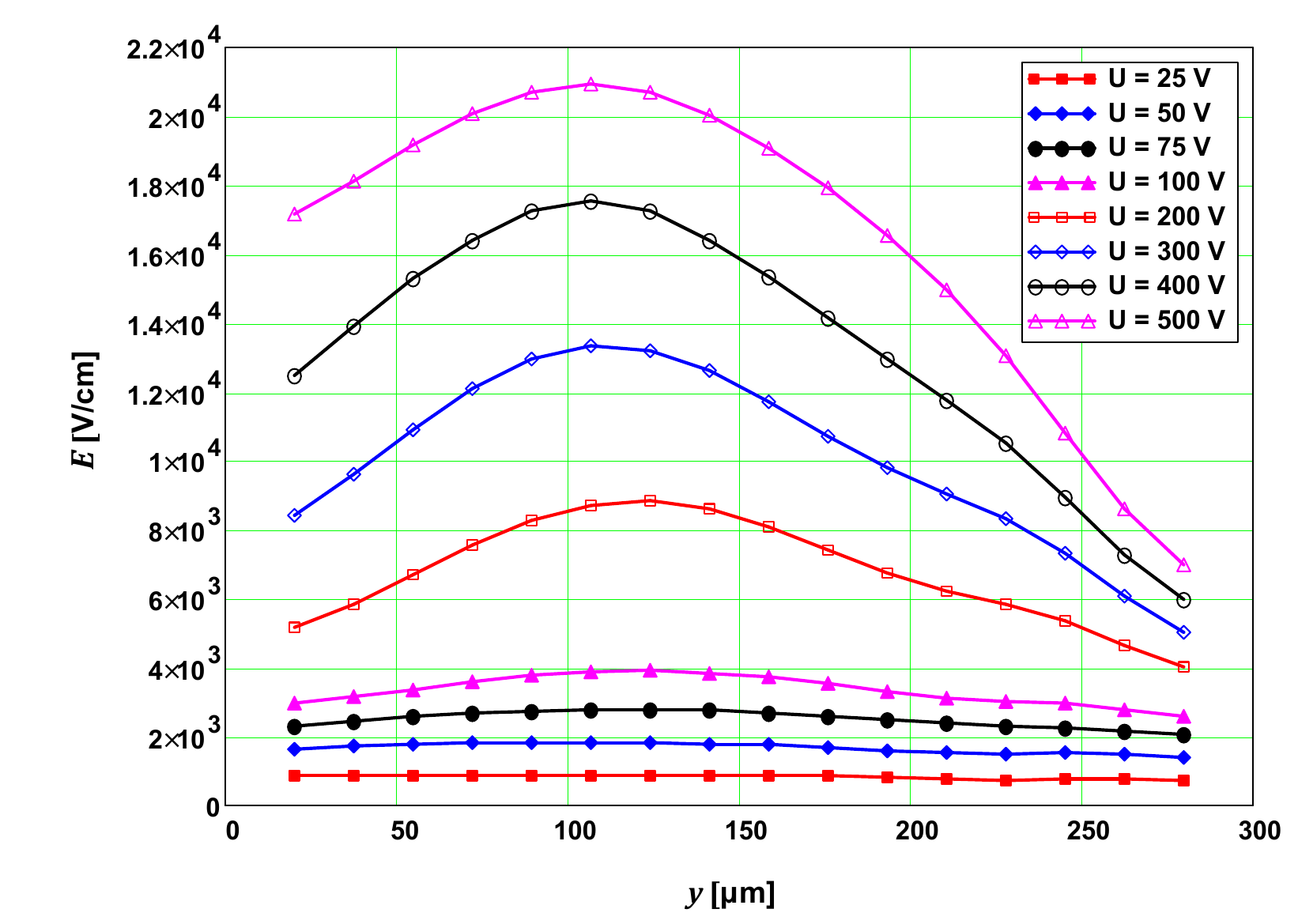}
   \caption{ }
    \label{fig:Efit-irr-forw-5E15}
   \end{subfigure}%
   \caption{
     Same as Fig.~\ref{fig:Fit-1E16-forw} for $\Phi _{eq} = 5 \times 10^{15}$~cm$^{-2}$.
    }
  \label{fig:Fit-5E15-forw}
 \end{figure}

\begin{figure}[!ht]
   \centering
   \begin{subfigure}[a]{0.5\textwidth}
    \includegraphics[width=\textwidth]{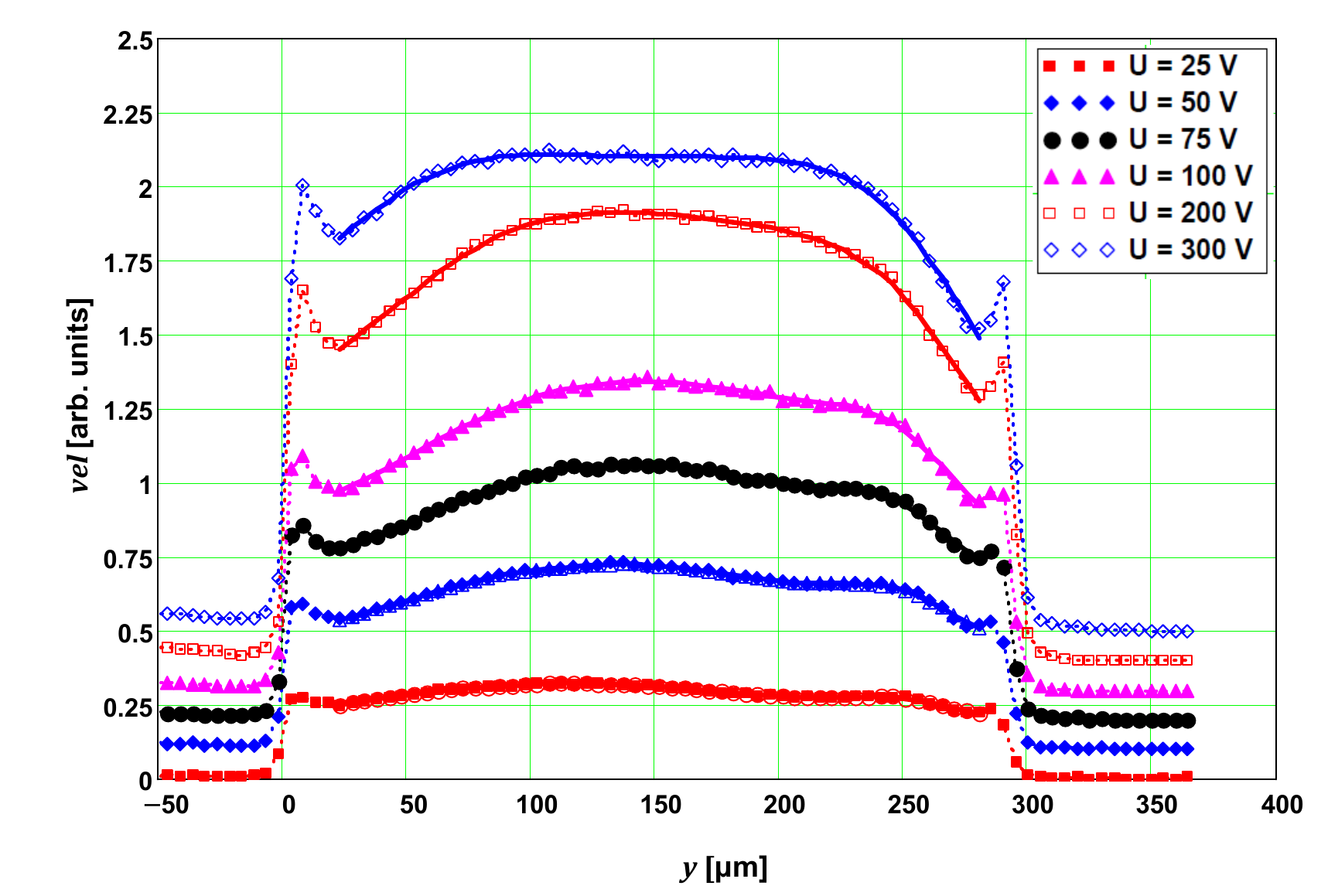}
    \caption{ }
    \label{fig:Vel-irr-forw-2E15}
   \end{subfigure}%
   \begin{subfigure}[a]{0.5\textwidth}
    \includegraphics[width=\textwidth]{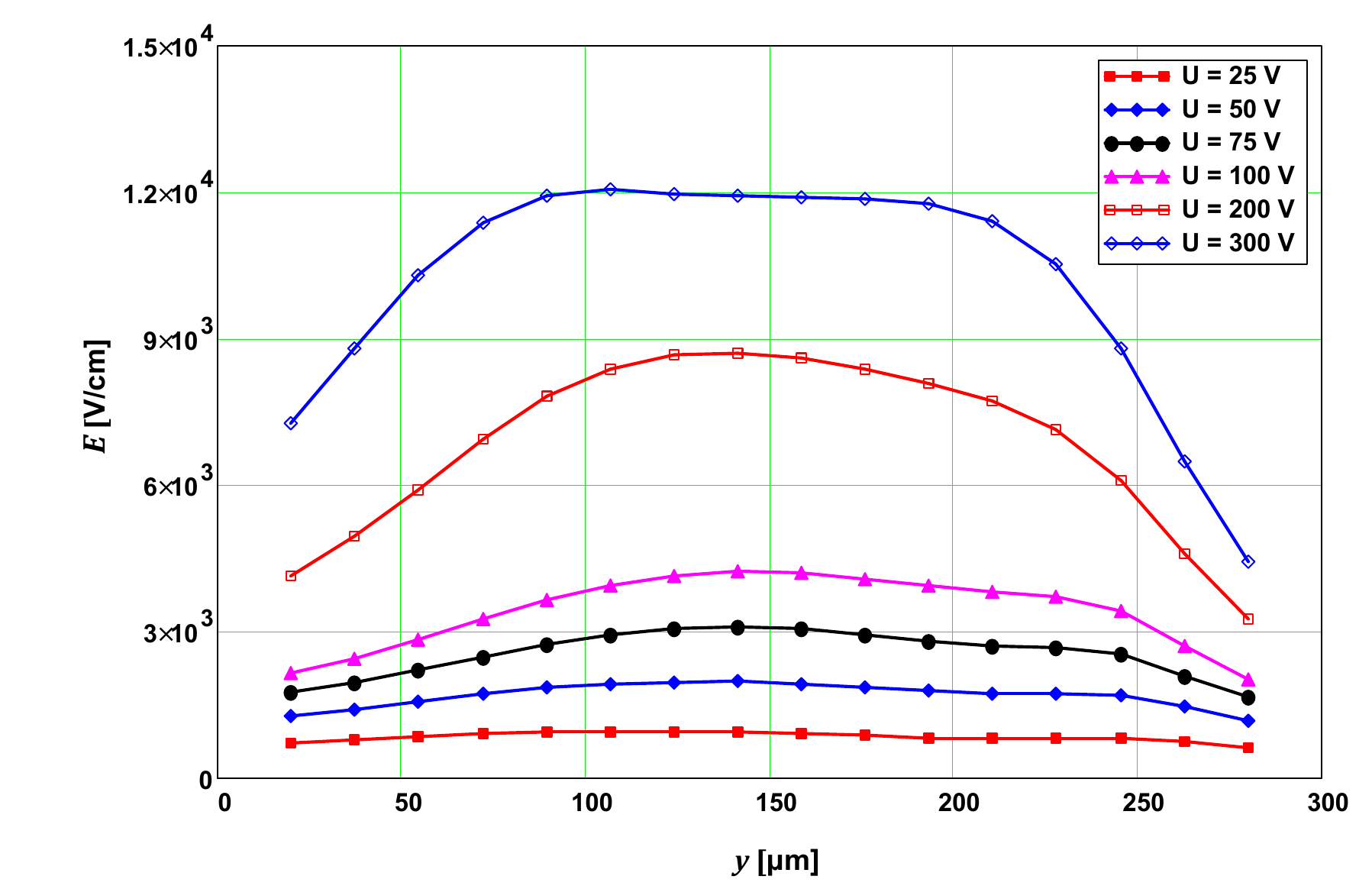}
   \caption{ }
    \label{fig:Efit-irr-forw-2E15}
   \end{subfigure}%
   \caption{
     Same as Fig.~\ref{fig:Fit-1E16-forw} for $\Phi _{eq} = 2 \times 10^{15}$~cm$^{-2}$.
    }
  \label{fig:Fit-2E15-forw}
 \end{figure}

 Figs.~\ref{fig:Fit-1E16-forw}, \ref{fig:Fit-5E15-forw} and \ref{fig:Fit-2E15-forw} show the results of the fits to the velocity profiles.
 For all $U$~values the velocity profiles are described by the fit within their statistical uncertainties.
 Varying the $y$~range of the fit has no significant effect on the $E$~values.
 At the lowest voltages $E$ is independent of $y$, as expected for an ohmic resistor.
 At higher voltages a transition to a $y$-dependent electric field is observed.
 The voltage at which the transition takes place increases with $\Phi _{eq}$.
 In Sect.~\ref{sect:Neff} these results are discussed, and compared to the ones for reverse bias.

 \subsection{Reverse bias}
  \label{sect:irr-rev}

  In this section the  analysis of the $vel$ data for reverse voltages and four neutron fluences from Ref.~\cite{Kramberger:2014} are presented.
  The neutron fluences are $\Phi _{eq} = (1, ~2, ~5 ~\mathrm{and} ~10) \times 10^{15}$~cm$^{-2}$, and the voltage range $U = 50 ~\mathrm{to} ~800$~V except for the  $5 \times 10^{15}$~cm$^{-2}$~data, where the maximum voltage is 750~V.
  The data were taken at $- 20 ~^\circ$C.
 \begin{figure}[!ht]
   \centering
   \begin{subfigure}[a]{0.5\textwidth}
    \includegraphics[width=\textwidth]{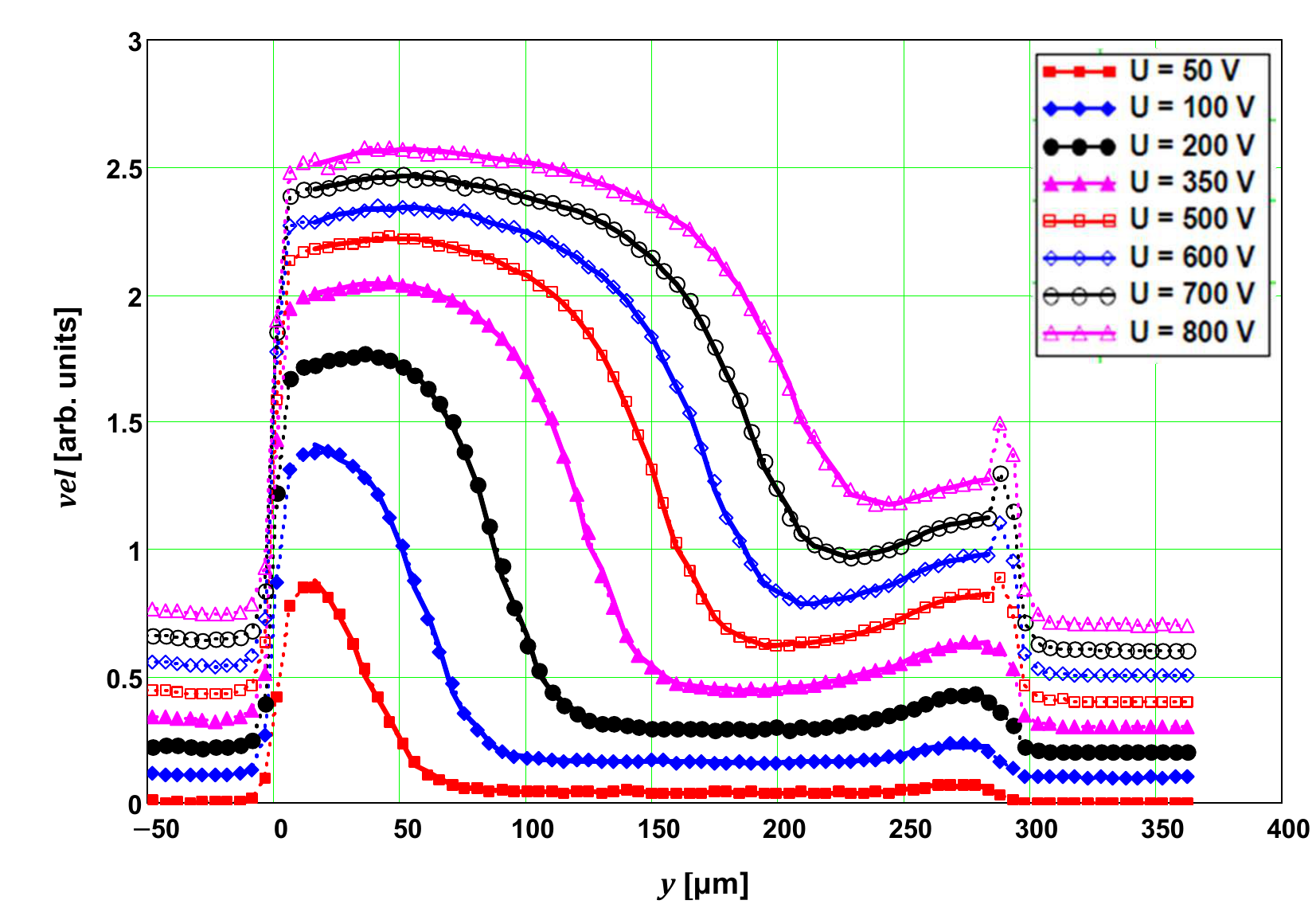}
    \caption{ }
    \label{fig:Vel-irr-rev-1E15}
   \end{subfigure}%
   \begin{subfigure}[a]{0.5\textwidth}
    \includegraphics[width=\textwidth]{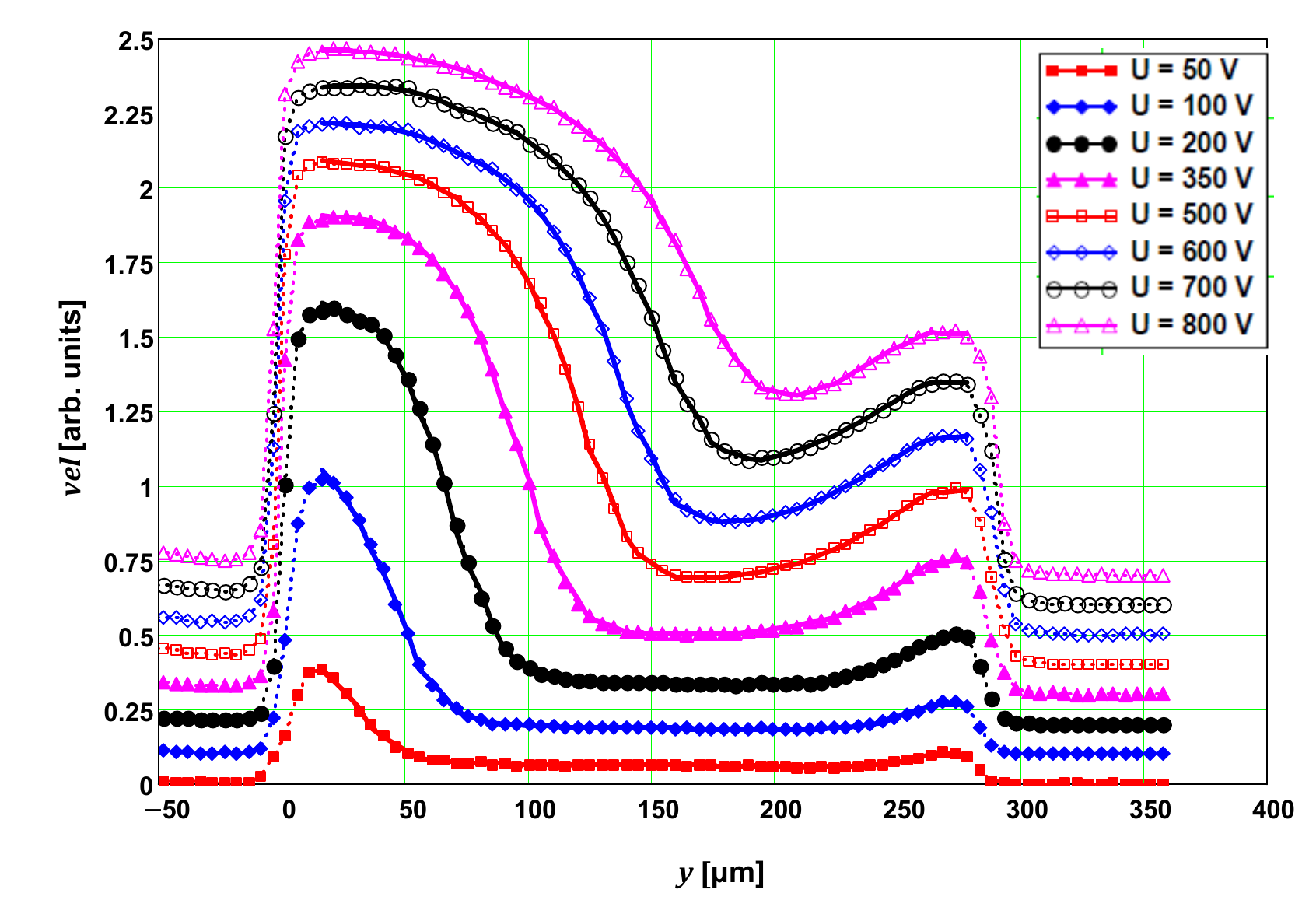}
   \caption{ }
    \label{fig:Efit-irr-rev-2E15}
   \end{subfigure}%
 \newline
   \begin{subfigure}[a]{0.5\textwidth}
    \includegraphics[width=\textwidth]{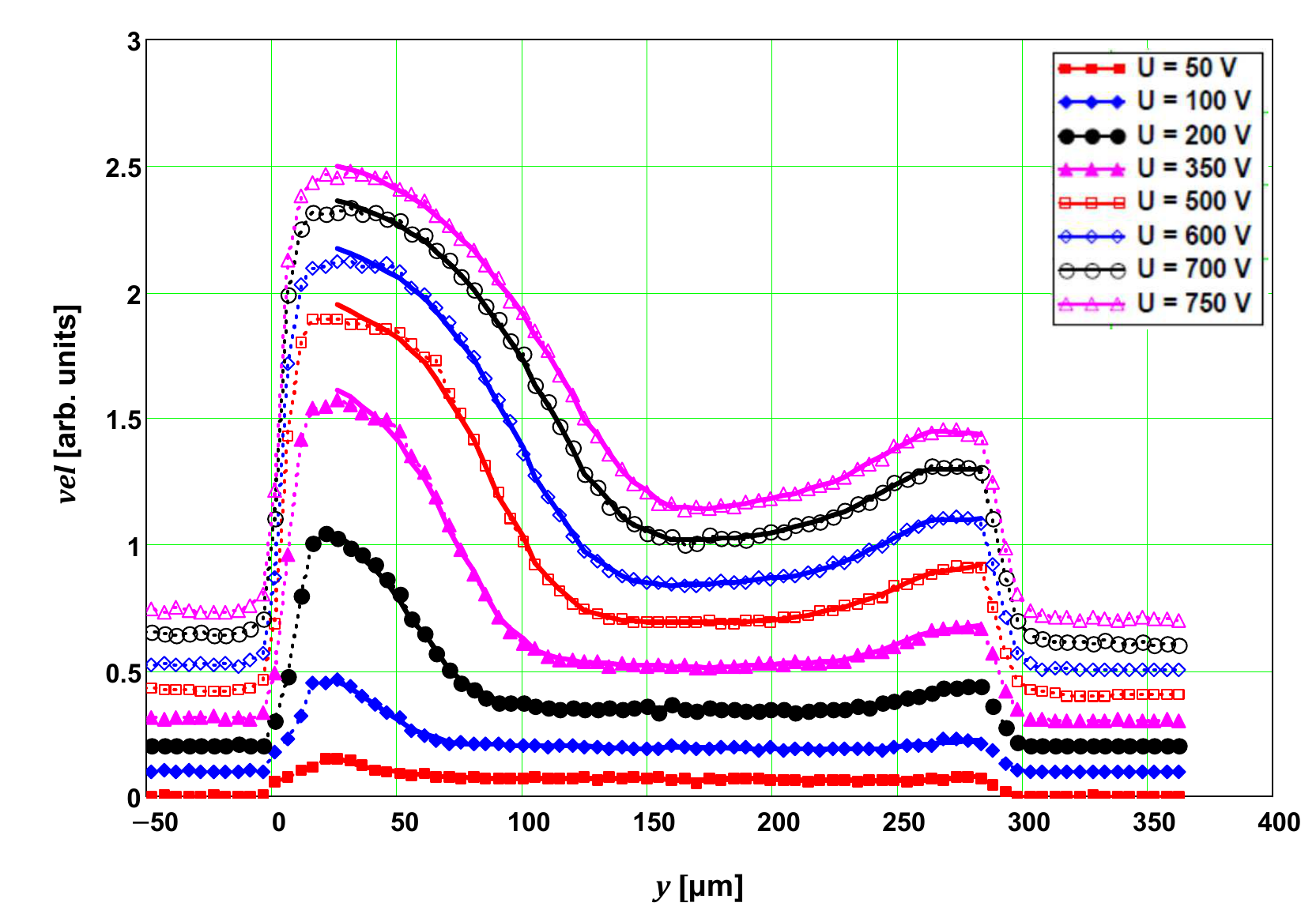}
    \caption{ }
    \label{fig:Vel-irr-rev-5E15}
   \end{subfigure}%
   \begin{subfigure}[a]{0.5\textwidth}
    \includegraphics[width=\textwidth]{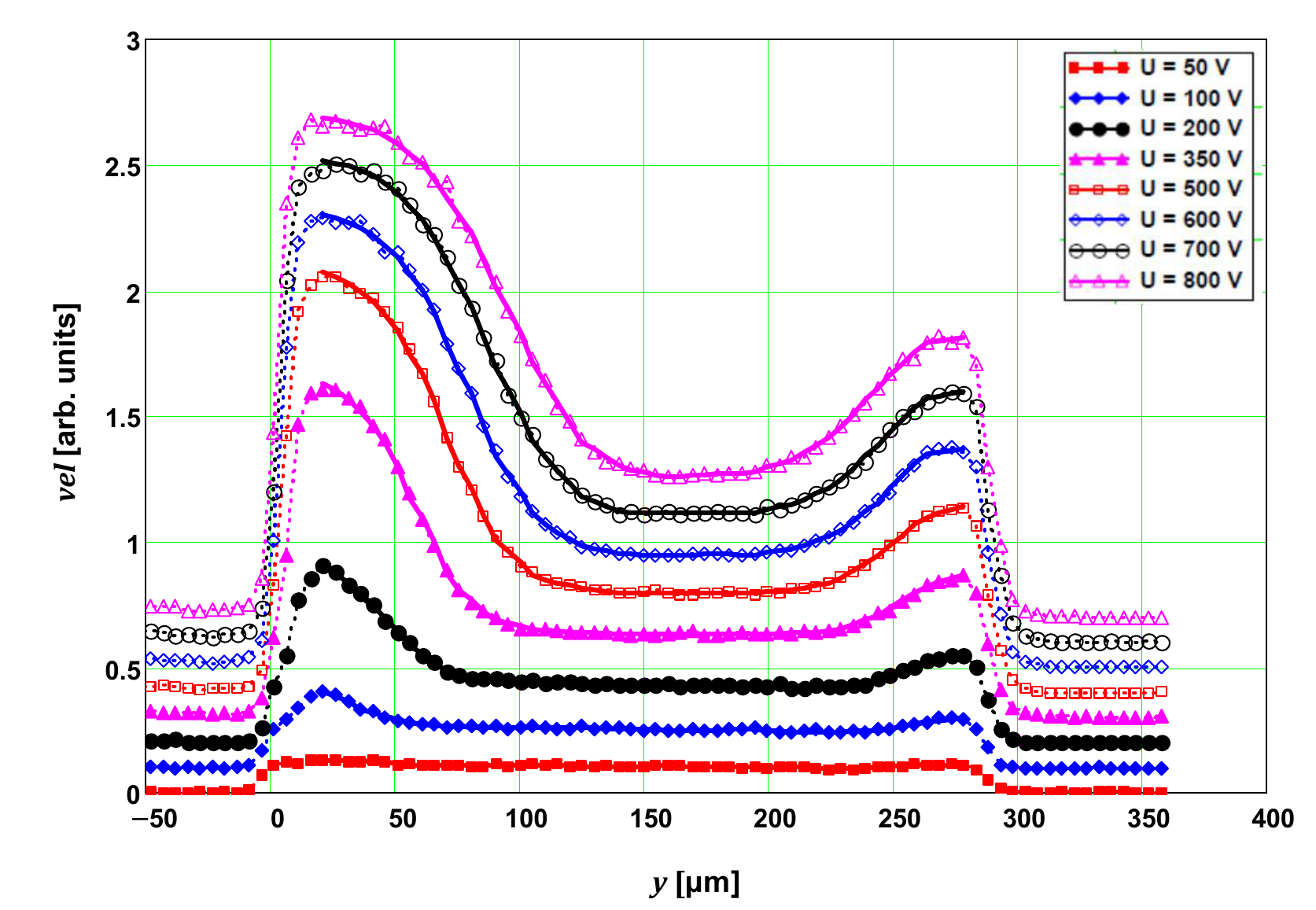}
   \caption{ }
    \label{fig:Vel-irr-rev-1E16}
   \end{subfigure}
   \caption{
     Fits to the velocity profiles for reverse bias for the $\Phi _{eq}$~values
      (a) $10^{15}$~cm$^{-2}$,
      (b) $2 \times 10^{15}$~cm$^{-2}$,
      (c) $5 \times 10^{15}$~cm$^{-2}$ and
      (d) $10^{16}$~cm$^{-2}$.
     For clarity, the individual data are shifted vertically by steps of 0.1.
     The symbols are the experimental data and the solid lines, which are hardly visible because of the good description of the data, are the results of the fit.
    }
  \label{fig:Vel-irr-rev}
 \end{figure}

 Fig.~\ref{fig:Vel-irr-rev} compares the measured $vel$~profiles to the fit results for  different reverse voltages and $\Phi _{eq}$.
 The fits are performed in the range $21~\upmu$m $\leqslant y \leqslant 278~\upmu$m.
 Changing the fit range makes little changes to the overall results, however if the lower $y$~value is reduced or the upper $y$~value increased, differences are observed in these $y$~ranges.
 Again, the model describes the data within their statistical uncertainties.

 \begin{figure}[!ht]
   \centering
   \begin{subfigure}[a]{0.5\textwidth}
    \includegraphics[width=\textwidth]{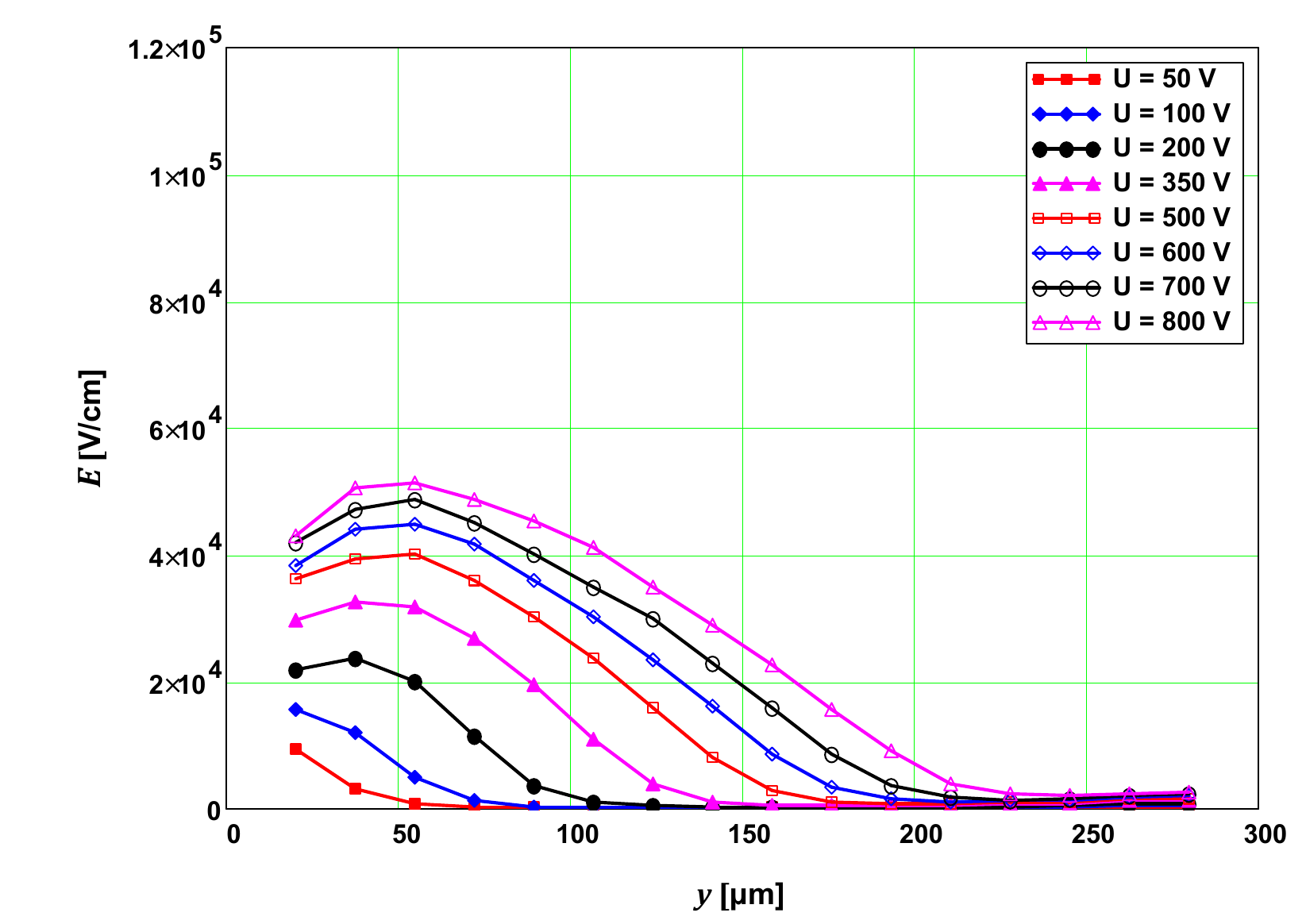}
    \caption{ }
    \label{fig:Efit-irr-rev-1E15}
   \end{subfigure}%
   \begin{subfigure}[a]{0.5\textwidth}
    \includegraphics[width=\textwidth]{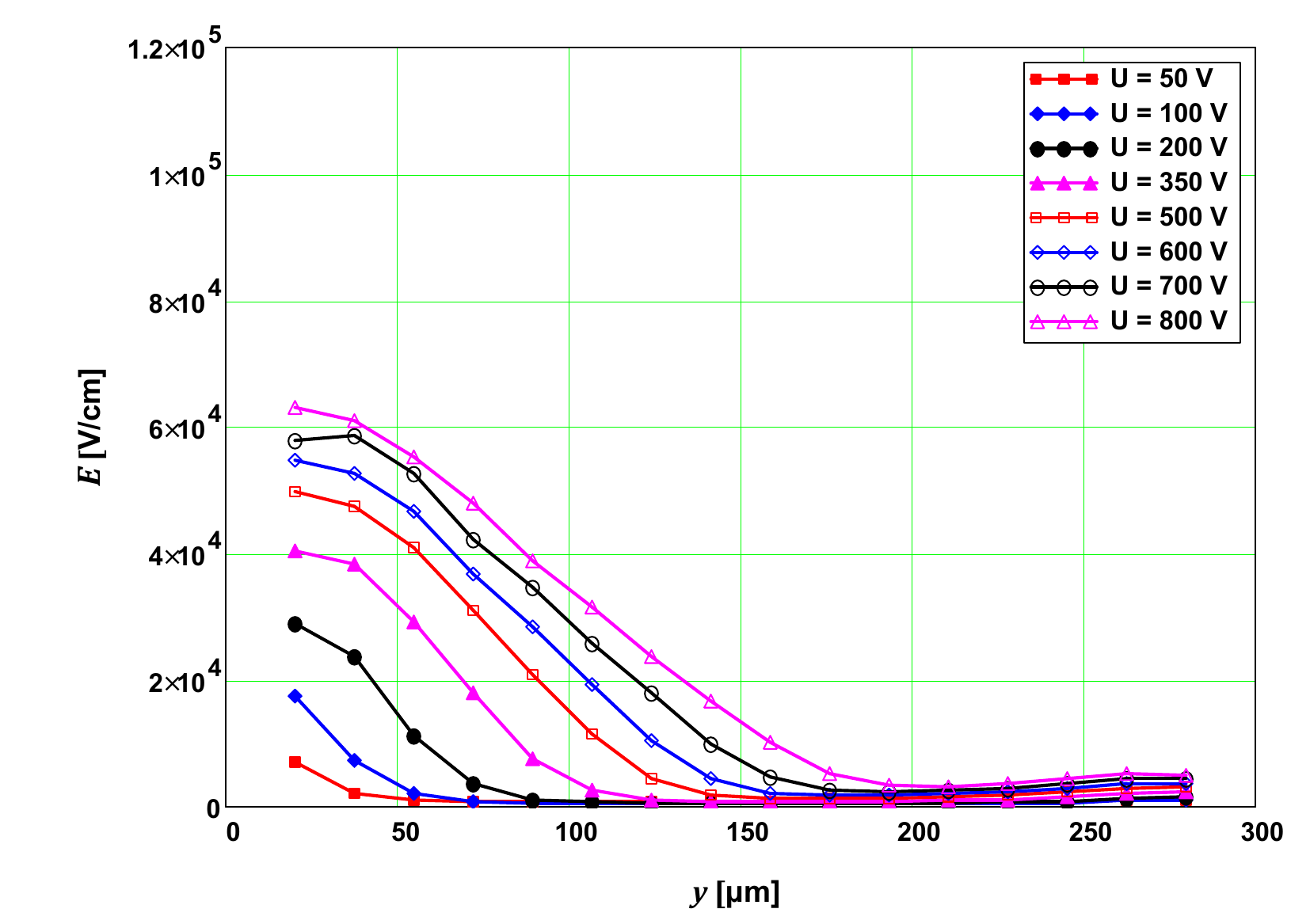}
   \caption{ }
    \label{fig:Efit-irr-rev-2E15}
   \end{subfigure}%
 \newline
   \begin{subfigure}[a]{0.5\textwidth}
    \includegraphics[width=\textwidth]{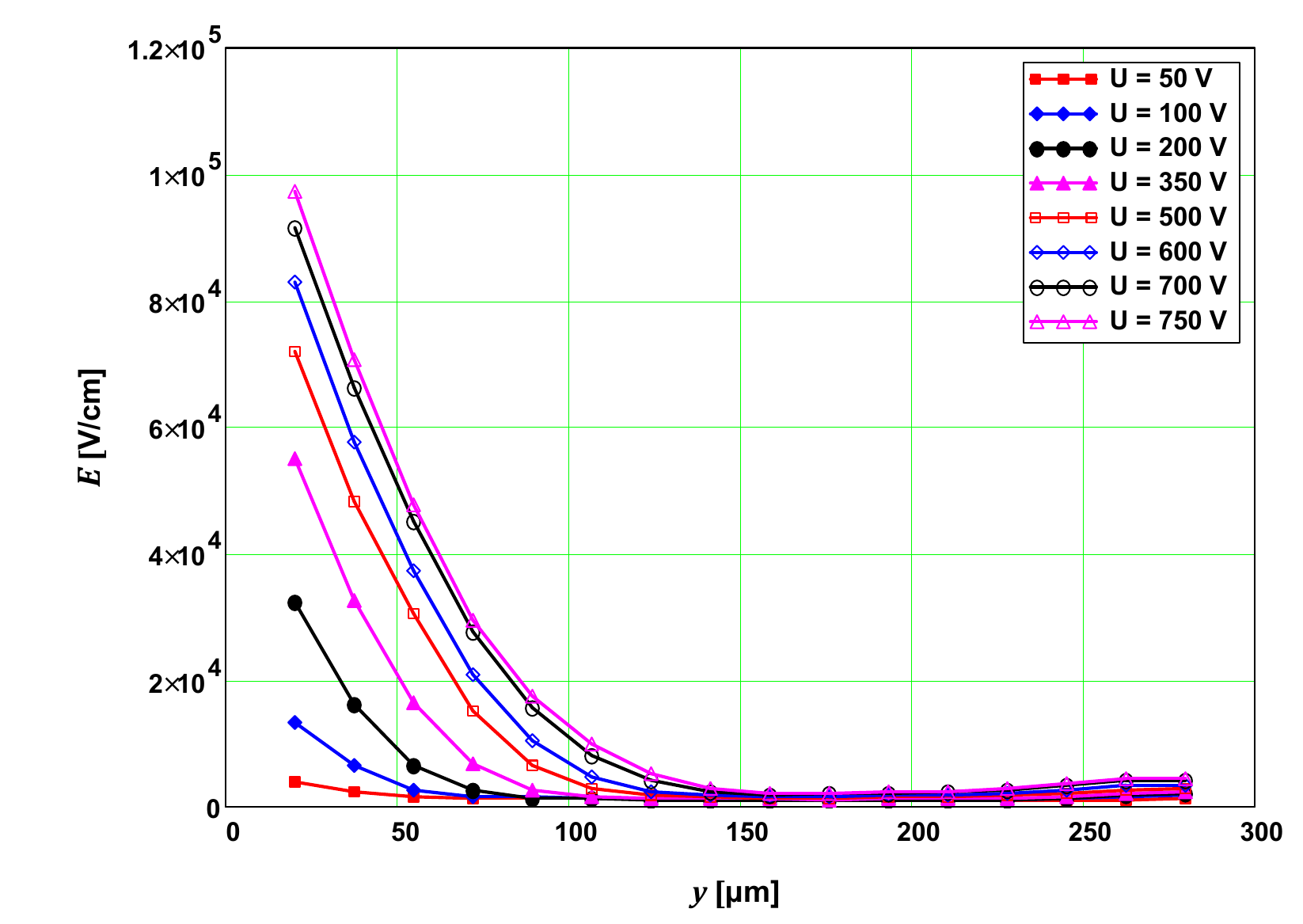}
    \caption{ }
    \label{fig:Efit-irr-rev-5E15}
   \end{subfigure}%
   \begin{subfigure}[a]{0.5\textwidth}
    \includegraphics[width=\textwidth]{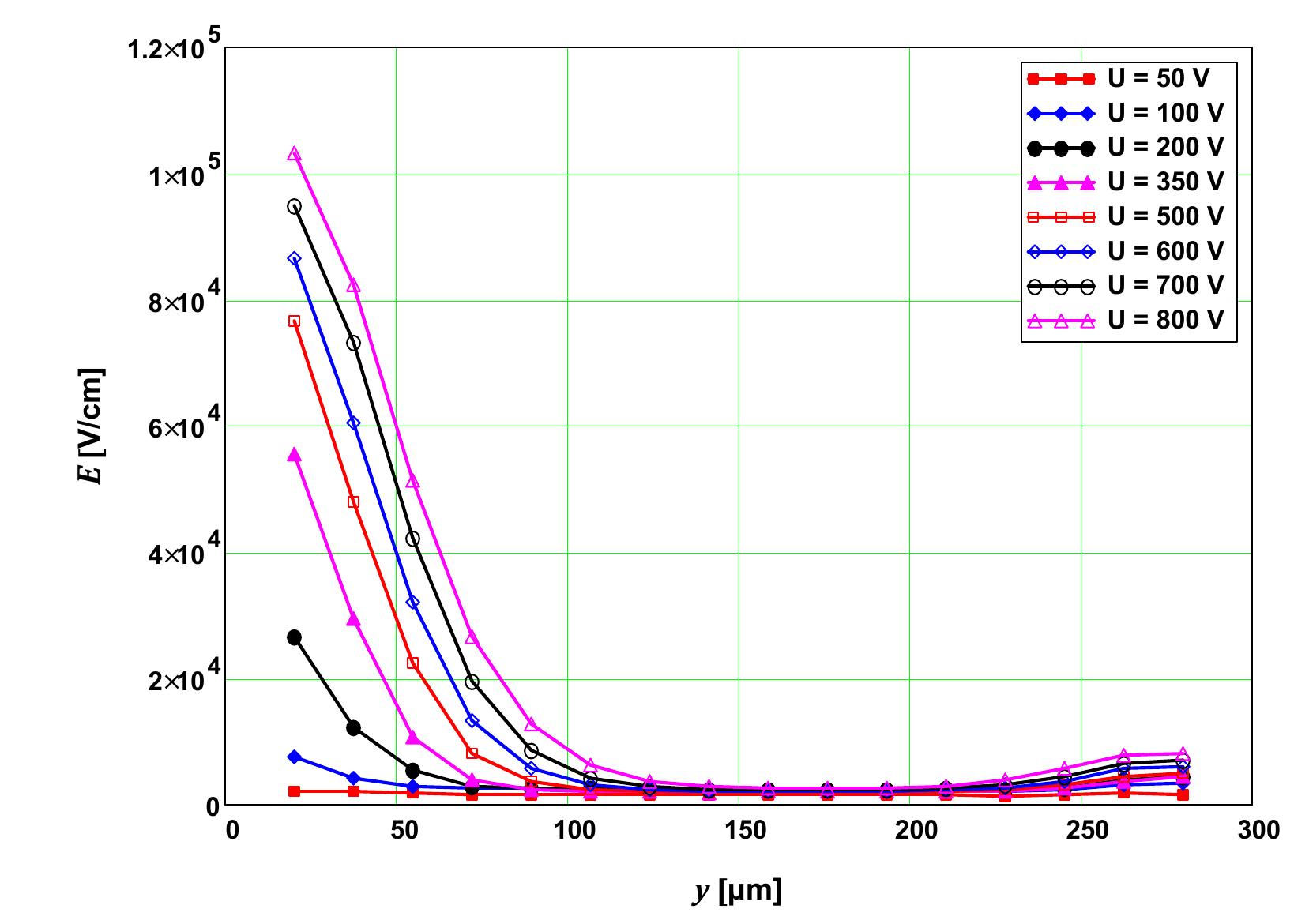}
   \caption{ }
    \label{fig:Efit-irr-rev-1E16}
   \end{subfigure}
   \caption{
    Electric fields determined by the fits to the velocity profiles for reverse bias for the $\Phi _{eq}$~values
      (a) $10^{15}$~cm$^{-2}$,
      (b) $2 \times 10^{15}$~cm$^{-2}$,
      (c) $5 \times 10^{15}$~cm$^{-2}$, and
      (d) $10^{16}$~cm$^{-2}$.
    }
  \label{fig:Efit-irr-rev}
 \end{figure}

 Fig.~\ref{fig:Efit-irr-rev} shows $E(y)$ obtained from the fits to  the  $vel$~profiles.
 The $y$~dependence $E(y)$ is opposite to what is observed for forward bias:
 maximal fields towards $y = 0$ and $y = d$ and  a low-field region in-between.
 The flattening of $E(y)$ towards $y=0$ at $\Phi_{eq} = 10^{15}$ and $2 \times 10^{15}$~cm$^{-2}$ is ascribed to the effects of the $x$-dependence of $E$ in strip sensors discussed in Sect.~\ref{sect:Model},  the response function of the readout discussed in Sect.~\ref{sect:AppendixB} and the finite width of the light beam.
 The values of $E$ in the central low-field region are compatible with the expectations from the field produced by the dark current with the resistivity of intrinsic silicon similar to the situation at low forward bias voltages.
 At $U = 50$~V and $\Phi_{eq} = 10^{16}$~cm$^{-2}$,  $E(y)$ is constant over the entire sensor.
 With increasing voltage and fluence $E$ increases at low $y$~values.
 At $U = 800$~V and $\Phi_{eq} = 10^{16}$~cm$^{-2}$, a value $E > 100$~kV/cm is reached, and the regime of charge multiplication is approached~\cite{Mandic:2009, Casse:2010, Lange:2010}.
 With increasing fluence and voltage the extension of the high-field region at low $y$ decreases.
 Also for $y$~values approaching $ y = d$ an increase of $E(y)$ is observed, which is called  \emph{double junction} in the literature~\cite{Eremin:2002}.
 However, compared to the electric field close to $y = 0$, the value towards $ y = d$ is much smaller and does not exceed 10~kV/cm.

   \subsection{Discussion of the results}
  \label{sect:Neff}

 This section illustrates how the results of the analysis of the precise edge-TCT data presented in this paper together with the dark current measurements provides an insight into the effects of radiation damage in silicon detectors.
 The formulae used for the following discussion can be found in standard textbooks on semiconductor devices~\cite{Lutz:1999, Sze:1981}.
 From the electric field values, $E_i$, the total charge carrier density, $N_{tot}(y)$, is obtained
 \begin{equation}\label{equ:Neff}
   N_{tot}(y) = \frac{\rho (y)} {q_0} =\frac{\mathrm{d}E(y)} {\mathrm{d}y} \cdot \frac{\varepsilon _{Si}} {q_0} = N^+(y) - N^-(y) +n_h(y) - n_e(y).
 \end{equation}
 $N_{tot}(y)$ can be positive and negative.
 The densities of fixed positive and negative space-charge carriers are denoted $N^+$ and $N^-$, respectively.
 They are given by the sum over the densities of the individual states in the silicon band gap multiplied with the probabilities that they are charged.
 Whereas the densities of the individual states for a given fluence $\Phi _{eq}$ are expected to be independent of position and voltage, the probabilities that the states are charged are not.
 They depend on the densities of free electrons, $n_e(y)$, and holes, $n_h(y)$, which are related to the current density, $\vec{j}$, and the electric field by
  \begin{equation}\label{equ:j}
   \vec{j} =  \vec{j}_e +  \vec{j}_h = \mu _ e \cdot (n_e \cdot q_0 \cdot \vec{E} + k \cdot T \cdot \vec{\nabla} n_e ) + \mu _h \cdot (n_h \cdot q_0 \cdot \vec{E} - k \cdot T \cdot \vec{\nabla} n_h ).
  \end{equation}
 For the steady state situation, $\partial n_e / \partial t = \partial n_h / \partial t = 0$ and from the current-continuity equation follows $\vec{\nabla} \vec{j} = 0$.
 From the additional model assumption that the electric field depends only on $y$ follows that $\vec{j}$ has only a $y$-component and that $j$ is constant in the entire detector.
 Its value is given by the dark current divided by the sensor area.
 Under the same assumptions the separate current-continuity equations for holes and electrons are
 \begin{equation}\label{equ:jeh}
   \mathrm{d}j_h/ \mathrm{d}y = q_0 \cdot (G_h - U_h) \hspace{5mm} \mathrm{and} \hspace{5mm}
   \mathrm{d}j_e/ \mathrm{d}y = - q_0 \cdot (G_e - U_e),
 \end{equation}
 with the generation rates, $G_h$ and $G_e$, and the recombination rates, $U_h$ and $U_e$, for holes and electrons, respectively.
 In the absence of an external source which generates electron-hole pairs in the detector, $G_h = G_e$ and $U_h = U_e$.
 In addition, $E(y)$ is constrained by $\int _0 ^d E(y)~\mathrm{d}y = U$, and the sign of $E(y)$ is the same in the entire detector.

\begin{figure}[!ht]
   \centering
   \begin{subfigure}[a]{0.5\textwidth}
    \includegraphics[width=\textwidth]{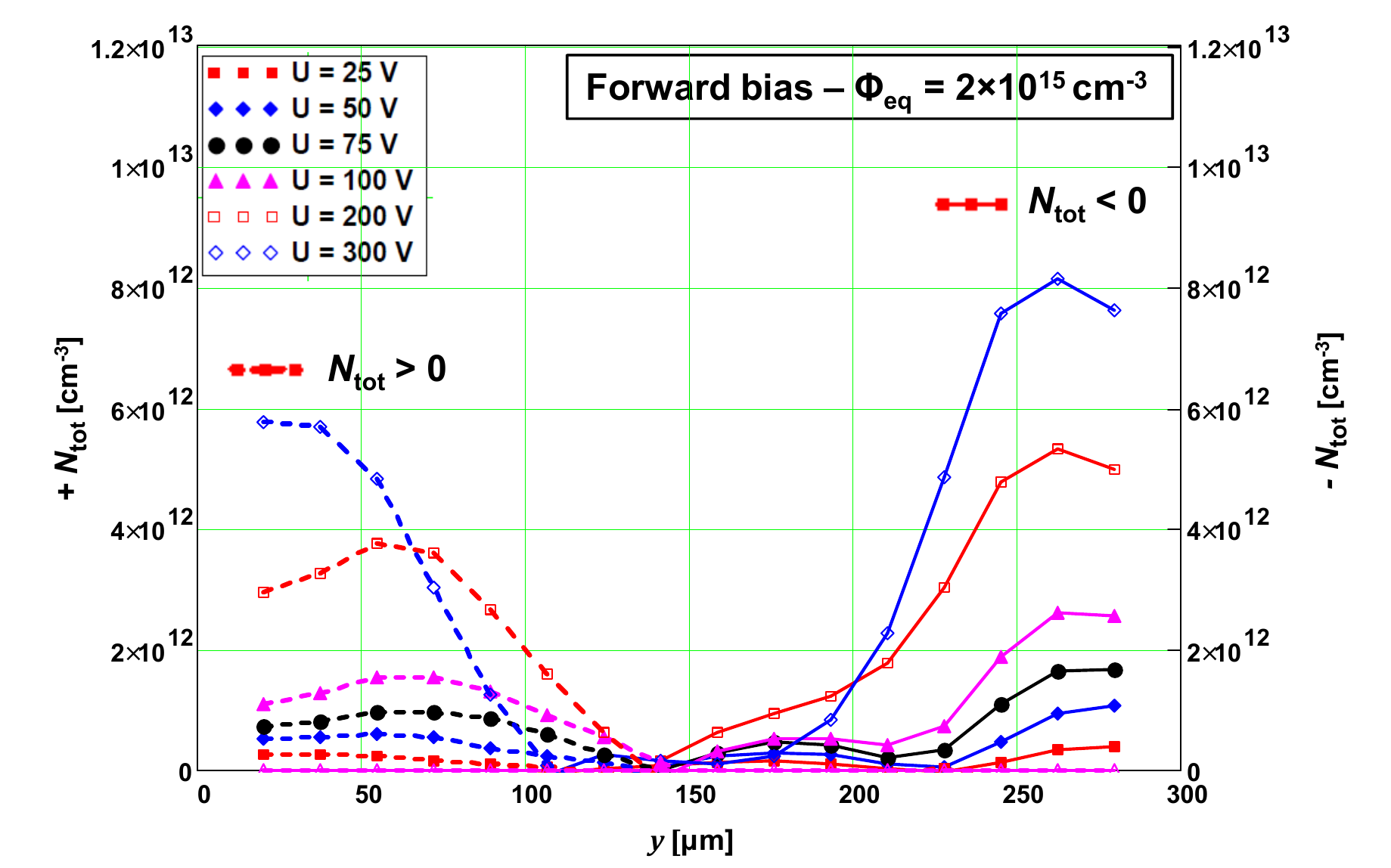}
    \caption{ }
    \label{fig:Neff-forw}
   \end{subfigure}%
   \begin{subfigure}[a]{0.5\textwidth}
    \includegraphics[width=\textwidth]{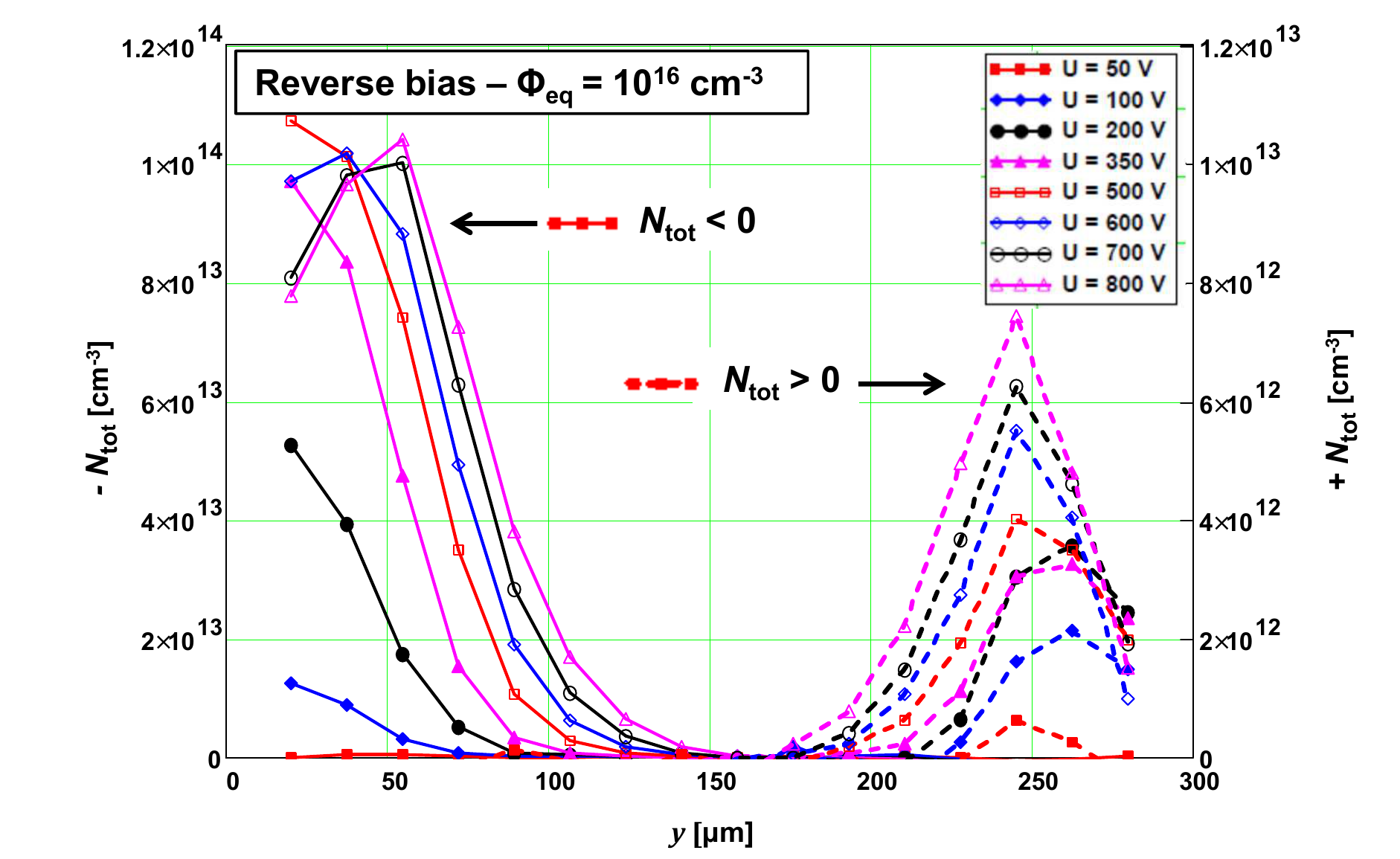}
    \caption{ }
    \label{fig:Neff-rev}
   \end{subfigure}%
   \caption{
    Total charge-carrier density distributions, $N_{tot}(y)$, obtained from the derivative $\mathrm{d}E / \mathrm{d}y$ for
      (a) forward bias and $\Phi _{eq} = 2 \times 10^{15}$~cm$^{-2}$, and
      (b) reverse bias and $\Phi _{eq} = 10^{16}$~cm$^{-2}$.
     The dashed lines show $ + N_{tot}$ and the solid lines $ - N_{tot}$.
     Note the different scales in (b) for $- N_{tot}$ (left) and $+ N_{tot}$ (right).
    }
  \label{fig:Neff}
 \end{figure}

 It has to be noted that taking derivatives can result in significant fluctuations.
 To avoid such problems, in Eq.~\ref{equ:Chisq} the number $n_E$ and the positions $y_i$, where the $E_i$~values are determined by the fit, as well as $w_{pen}$ have to be optimised for every voltage and fluence.
 Such a study has not been done, and therefore just two examples, one for forward and one for reverse bias are shown, for which the fluctuations after differentiation appear acceptable.
 Fig.~\ref{fig:Neff} shows the two examples for $N_{tot}(y)$ for selected bias voltages $U$.

 For forward bias, Fig.~\ref{fig:Neff-forw} shows that $N_{tot}(y)$ is positive at low $y$ and negative at high $y$.
 This is expected, because electrons move in the $+y$-direction and holes in the $-y$-direction, resulting in an excess of holes at low $y$ and an excess of electrons at high $y$, which changes the occupation of the states in the silicon band gap accordingly.
 For reverse bias the situation is opposite: holes move in the $+y$-direction and electrons in the $-y$-direction, and the sign of $N_{tot}(y)$ is reversed, as can be seen in Fig.~\ref{fig:Neff-rev}.
 In both cases the sign of $N_{tot}$ changes approximately in the middle of the detector, where extended low $|N_{tot}|$-regions are observed.
 Towards the electrodes $|N_{tot}|$ has maxima.

 First the results for forward bias and low voltages ($U \lesssim 50$~V for $\Phi _{eq} = 2 \times 10^{15}$, and $\lesssim 100$~V for $10^{16}$~cm$^{-2}$) are discussed.
 Electrons are injected through the forward-biased $n^+p$-junction at $y = 0$.
 They drift in the $+y$-direction and along their path recombine until the recombination-generation equilibrium is reached.
 Similarly, holes are injected through the $p^+p$-junction at $y = d$, recombine on their path in the $-y$-direction until the recombination-generation equilibrium is reached.
 In these transition regions, a small increase of $|N _{tot}|$, typically below $10 ^{12}$, is observed.
 In the central equilibrium region, the density of electrons is expected to be $n_e = \sqrt{ \mu _h (0) / \mu _e (0)} \cdot n_i$, and the density of holes  $n_h = \sqrt{ \mu _e (0) / \mu _h (0)} \cdot n_i$~\cite{Schwandt:2019}
 The low-field mobilities are $\mu _e (0)$ and $\mu _h (0)$, and the intrinsic charge carrier density $n_i \approx 1.2 \times 10 ^8$~cm$^{-3}$ at $-20~^\circ$C.
 Fig.~\ref{fig:Neff-forw} shows that in this region $|N_{tot}| < 10^{11}$~cm$^{-3}$, and the electric field is the result of the ohmic resistance of the current given by Eq.~\ref{equ:j}: $E = j / \big(q_0 \cdot (n_e \cdot \mu _e + n_h \cdot \mu _h) \big)$.
 For low voltages, the electric field shown in Fig.~\ref{fig:Efit-irr-forw-2E15} agrees within $\pm 20$~\% with this prediction.
 At small $y$-values, where $n_e \gg n_i$ the value of $E$ has to decrease in order to satisfy  the requirement of a constant $j$.
 The situation is analogous for $y$ approaching $d$, where holes are injected through the $p^+p$-junction and $n_h \gg n_i$.
 Fig.~\ref{fig:Efit-irr-forw-2E15} shows that this decrease is actually observed at both sides of the detector.
 The slower increase of $E (y)$ for low $y$-values compared to the decrease at high~$y$-values is expected because $\mu _e > \mu _h$.
 The observation from Figs.~\ref{fig:Efit-irr-forw-1E16}, \ref{fig:Efit-irr-forw-5E15} and \ref{fig:Efit-irr-forw-2E15} that at higher $\Phi _{eq}$-values the equilibrium condition is reached at shorter distances and also extends to higher voltages, is explained by the higher density of radiation-induced states and therefore higher recombination rates.
 For higher forward voltages, regions with positive $N_{tot}$ develop for small values of $y$, and with negative  $N_{tot}$ at larger $y$-values.
 In-between, $N_{tot}$ remains low, and the electric field is high and approximately constant with a value, well in excess of $U / d$, which can be seen in Fig.~\ref{fig:Efit-irr-forw-2E15}.
 This region is similar to a depletion region with low doping concentration.

 The situation for reverse bias, where the electric field points in the $+y$-direction, is very different.
 The reverse-biased $n^+p$-junction prevents the injection of holes at $y = 0$, and the $p^+p$-junction at $y = d$ the injection of electrons, and the dark current is caused by the generation of free charge carriers.
 As can be seen in Fig.~\ref{fig:Neff-rev} for reverse voltages as low as 50~V, $|N_{tot}| \lesssim 10^{11}$~cm$^{-3}$, and the electric field is approximately uniform and given by the ohmic resistivity of the damaged silicon and the current density (see Fig.~\ref{fig:Efit-irr-rev-1E16}).
 Thus the situation is similar to the one for low forward voltages.
 With increasing reverse voltage, regions of high $|N_{tot}|$ develop towards both electrodes.
 Their sign is negative for small $y$, where the dark current is dominated by electrons, and positive at high $y$-values, where it is dominated by holes.
 As can be seen from Fig.~\ref{fig:Neff-rev} the value of $|N_{tot}|$ at low $y$ is approximately a factor 20 higher than at high $y$.
 It can also be seen that with increasing voltage,  the two boundaries of the high-$|N_{tot}|$ region move towards the centre of the detector and the low-field region shrinks.
 However, for  $\Phi _{eq} = 10^{16}$~cm$^{-2}$ even at a voltage as high as 800~V, an ohmic low-field region remains in the centre.

 In the region of high $|N_{tot}|$ the contribution of $n_e$ and $n_h$ is negligible, and $N_{tot} \approx N^+ - N^-$.
 As for a given $\Phi_{eq}$ the density of radiation-induced levels does not depend on position nor on applied voltage, the value of $N_{tot}$ directly reflects the fraction of these levels which are in a charged state.
 It is also noted that typical values for the sum of the introduction rates of all defects is of order 1~cm$^{-1}$~\cite{Kramberger:2019} resulting in a density of approximately $10^{16}$~cm$^{-3}$ defect states for $\Phi _{eq} = 10^{16}$~cm$^{-2}$.
 Thus the observation of $|N_{tot}| \approx 10^{14}$~cm$^{-3}$ indicates that only a small fraction these states are charged.
 It should also be noted that, as discussed in Refs.~\cite{Pintilie:2009, Kramberger:2019}, the introduction rates for the different damage states and the removal of dopants depend on the type of particles used for the irradiation, and the results for irradiation with high- and low-energy charged hadrons can well be quite different.

Comparing $E(y)$ for forward and reverse bias reveals that for forward bias the region of high electric fields extends over a significantly larger region of the detector.
This explains why for high $\Phi _{eq}$ the measured charge collection for forward bias is higher than for reverse bias~\cite{Scharf:2018, Chilingarov:1997, Mandic:2004}.


  \section{Summary and conclusions}
  \label{sect:Summary}

 A straight-forward method is proposed and used to determine the position-dependent electric field and the total charge density of radiation-damaged silicon detectors from velocity profiles measured by edge-TCT (Transient Current Technique).
 The velocity profiles are extracted from the initial slopes of the current transients from charges produced by the light of a near-infrared laser beam injected into a strip sensor parallel to its surface.
 Simulations of the current transients for a non-irradiated sensor convolved with a simplified response function are used to investigate the validity and the limitations of the method.
 The main difficulties arise from the finite bandwidth of the readout and the poor knowledge of the electric field in the region close to sensor surface at which the strip electrodes are located.
 For the strip sensors investigated the effect is particularly important because of the small ratio of implant-width to strip-pitch of $20~\upmu$m to $100~\upmu$m.

 The method is first applied to data from a non-irradiated $n^+p$ sensor of $300~\upmu$m thickness for reverse voltages below and above the full-depletion voltage.
 The model provides a description of the measured velocity profiles within their statistical uncertainty and the electric fields approximately agree with the expectations of a non-irradiated pad sensor.
 In particular the expected voltage dependence of the depletion depth is well reproduced.

 Next the velocity profiles of $300~\upmu$m thick $n^+p$ strip sensors irradiated by reactor neutrons to neutron-equivalent fluences $\Phi _{eq} = (2,~5~\mathrm{and}~10)\times 10^{15}$~cm$^{-2}$ for forward voltages between 25~V and up to 500~V are analysed.
 Again, the model describes the experimental data within their statistical uncertainties.
 In agreement with expectations it is found that at low voltages the electric field in the sensor is the result of the ohmic voltage drop of the dark current in silicon with intrinsic resistivity.
 The intrinsic resistivity is the result of the generation-recombination equilibrium reached in highly radiation-damaged silicon at low electric fields.
 At higher voltages a high field region develops in the centre of the detector, which extends over most of the sensor depth.
 This explains the observation that above a certain radiation dose the charge collection efficiency is higher for forward than for reverse bias.

 Finally the method is used to analyse the velocity profiles for reverse voltages between 50~V and 800~V of a $300~\upmu$m thick $n^+p$ strip sensor irradiated by reactor neutrons to $\Phi _{eq} = (1,~2,~5~\mathrm{and}~10)\times 10^{15}$~cm$^{-2}$.
 Again, the model describes the experimental data within their statistical uncertainties.
 The electric field  has its maximum at the $n^+p$-junctions of the readout strips.
 The maximum field increases with reverse voltage, and also with $\Phi _{eq}$ at  higher voltages.
 At 800~V its value exceeds 100~kV/cm, which is close to the onset of charge multiplication for electrons.
 The extension of the high-field region decreases with increasing fluence.
 At 800~V and $\Phi _{eq} = 10^{16}$~cm$^{-2}$ the high-field region is less than $100~\upmu$m deep, which explains the observation that for high radiation damage the charge collection of $150~\upmu$m and $300~\upmu$m thick sensors is similar, i.~e. that the charge collection efficiency of thin silicon sensors is superior.
 With increasing $\Phi _{eq}$ and increasing reverse voltage an increase of the electric field at the $p^+p$ back side of the sensor is observed -- a phenomenon known as double junction.
 However, the maximum field is below 10~kV/cm, thus more than an order of magnitude smaller than at the $n^+p$-junction.
 From the derivative of the electric field with respect to the depth in the sensor the position-dependent total charge density, $N_{tot}$, is estimated and its sign determined.
 For $\Phi _{eq} = 10^{16}$~cm$^{-2}$ and reverse voltages above 300~V  $N_{tot} \approx 10^{14}$~cm$^{-3}$ and the charge sign is negative.
 In the region towards the $p^+p$-junction $N_{tot}$ increases steadily with increasing reverse voltage and a value of $N_{tot} \approx 7 \times 10^{12}$~cm$^{-3}$ is reached at 800~V.
 The charge sign is positive in this region.

 It is concluded that the method proposed in this paper successfully describes the velocity profiles from edge-TCT measurements and allows to extract the position-dependent electric field and the charge density in highly radiation-damaged silicon sensors.
 Although the main emphasis of the paper is on the method and its limitations, its application to sensors damaged by reactor neutrons already provides quite some insight into the changes of the detector properties due to radiation damage.
 Such information can also be used to verify the results of TCAD simulations based on radiation-induced states in the silicon band gap.
 As a next step the method should be used to analyse edge-TCT data for sensors irradiated by charged particles of different energies to understand the impact on the sensor performance of the experimentally observed differences in introduction rates of the different damage states.
 As one of the uncertainties in the analysis is the small ratio of strip implant to strip pitch, measurements with sensors with a smaller inter-strip distance are desirable.

 An extension of the method to analyse the entire current transient and not only its initial rise appears feasible.
 At least in principle, this extension would allow to determine the position-dependent lifetimes of holes and electrons in addition to the position-dependent electric field.
 Such information would enable a complete description of the performance of highly radiation-damaged sensors and thus be an important element in the simulation and data analysis of silicon sensors in high-radiation environments.

  \section{Appendix A: Effective weighting field for edge-TCT}
   \label{sect:AppendixA}

 In Sect.~\ref{sect:Model} it is shown that for a charge distribution uniformly distributed in $x$ (see Fig.\ref{fig:Sensor}) the effective weighting field required to evaluate the edge-TCT data, $\overrightarrow {\mathcal{E}_w} = \hat{e}_y/d$, the weighting field of a fully depleted pad detector.
 In this section numerical calculations are used to calculate $\overrightarrow {\mathcal{E}_w}(y)$ for a charge distribution $\propto e^{- x / \lambda_{abs}}$, where $\lambda _{abs}$ is the light-absorption length.
 For the wavelength $\lambda = 1064$~nm,  the light used for the measurements, $\lambda _{abs} \approx 1.7$~mm for non-irradiated silicon at a temperature of
 $ - 20~^\circ$C~\cite{Macfarlane:1958}.
 At this wavelength $\lambda _{abs} $ decreases with temperature.
 In addition, a decrease by approximately 15~\%  after irradiation to $\Phi _{eq} = 10^{16}$~cm$^{-2}$ has been observed at $ +20~^\circ$C~\cite{Scharf:2019}.

 Using the program of Ref.~\cite{Meeker:2018} the weighting field of the strip sensor has been simulated.
 The center of the central strip, to which the potential of +1~V was applied, is positioned at $x = y =0$ (see Fig.~\ref{fig:Sensor}).
 A total of 13 strips, 6 for negative and 6 for positive $x$ were simulated.
 The potential of the  strips excluding the central one, of the rear electrode at $y=d$ and of the silicon at the centers of the outer strips at $ x = \pm ~600~\upmu$m were set to zero.
 For the interface in-between the strips at $y = 0$ symmetric boundary conditions were assumed.
 The simulated potential has a similar shape as the potential of  half of the potential shown on the right-hand side of Fig.~\ref{fig:Potential} for the Neumann boundary conditions on the SiO$_2$~surface.
 Both $x$ and $y$-components of the weighting field, $\overrightarrow {{E}_w}$, vary strongly with $x$ and $y$.

 The effective weighting field is obtained from the integral
 $\overrightarrow {\mathcal{E}_w} = \big( \int \overrightarrow {{E}_w} (\vec{r}) \cdot e^{- x / \lambda_{abs}}~\mathrm{d}x  \big) / p $.
 As expected, for $\lambda _{abs} \gg p$ the $x$-component of $\overrightarrow {\mathcal{E}_w}$ is zero and the $y$-component is independent of $y$.
 However, the value is $\approx 30$~cm$^{-1}$ instead of $1/d = 33 $~cm$^{-1}$.
 The reason is that the region outside of the $\pm ~600~\upmu$m simulated contributes $\approx 10$~\% to $\mathcal{E} _w$.
 It is concluded that edge-TCT measurements should not be taken for strips too close to the edge of the sensor, and that several of the strips adjacent to the readout strip have to be connected to ground, ideally with the impedance of the readout electronics.

  \begin{figure}[!ht]
   \centering
   \begin{subfigure}[a]{0.5\textwidth}
    \includegraphics[width=\textwidth]{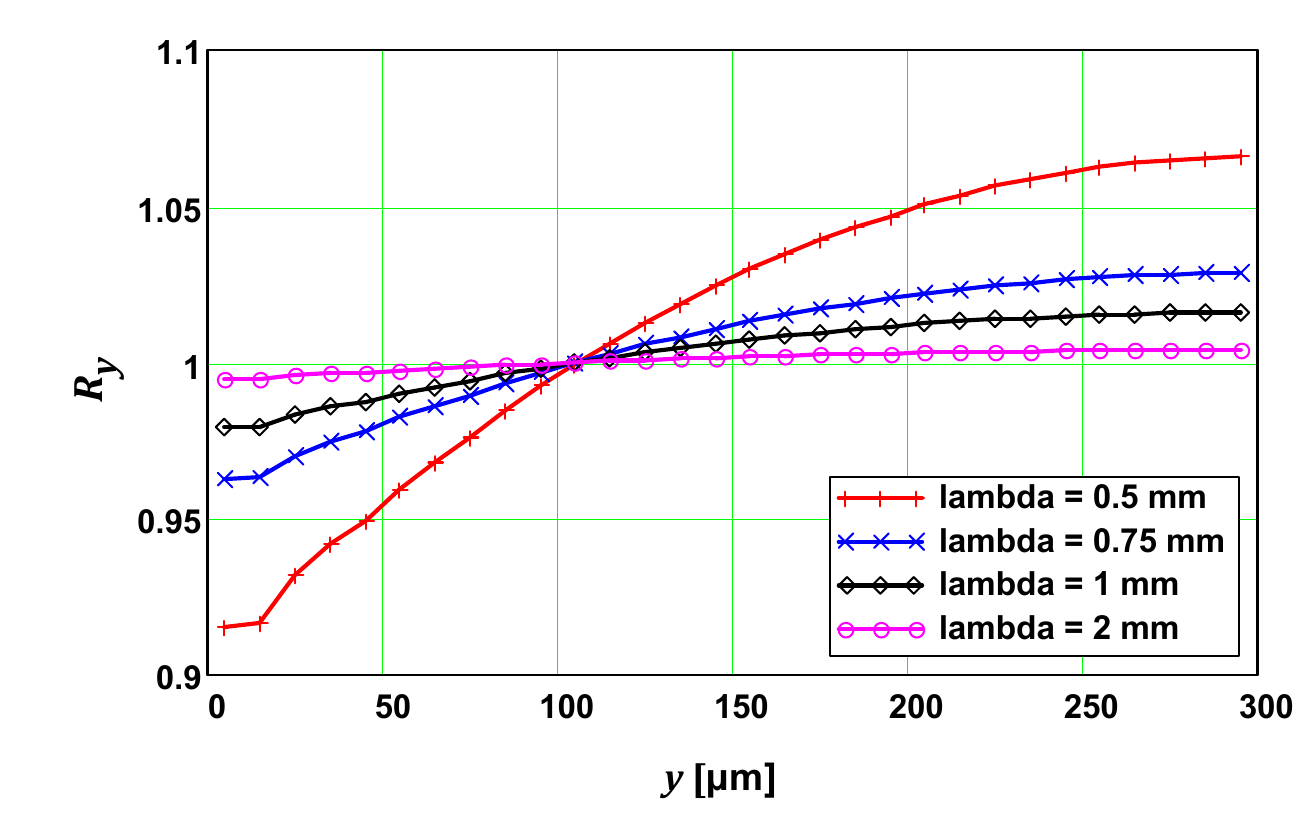}
    \caption{ }
    \label{fig:Rlambda}
   \end{subfigure}%
   \begin{subfigure}[a]{0.5\textwidth}
    \includegraphics[width=\textwidth]{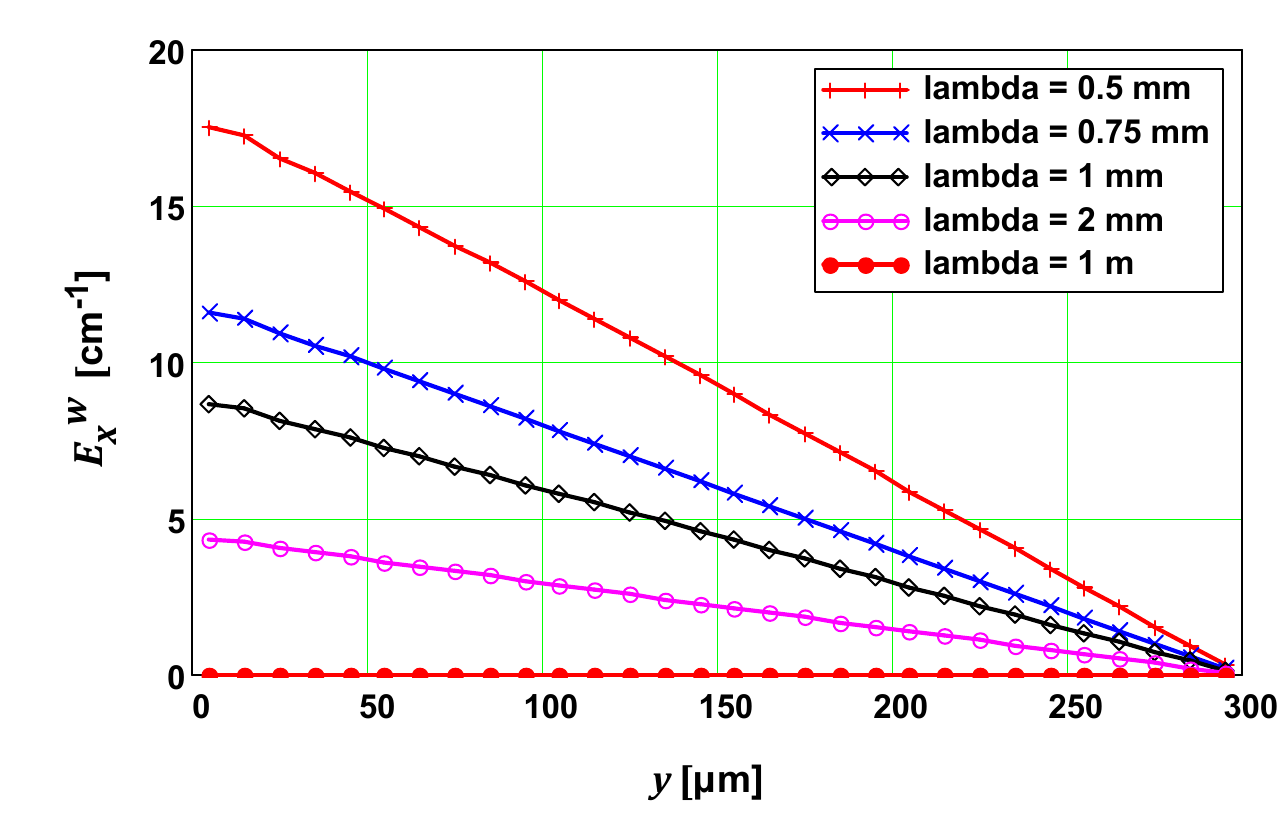}
   \caption{ }
    \label{fig:Ex-weight}
   \end{subfigure}%
   ~
   \caption{Influence of a finite light absorption length $\lambda _{abs}$ on $y$-dependence of the effective weighting field $ {\mathcal{\vec{E}} }_w$
    for a strip sensor with the thickness $d = 300~\upmu$m and the strip pitch $p = 100~\upmu$m.
    (a) $R_y$, the ratio of the $y$-component of $\mathcal{\vec{E}}_w$ for different values of $\lambda _{abs} $ to the value of $\mathcal{E}_w$ for $\lambda_{abs} = 1$~m.
    (b) $x$-component of $\mathcal{\vec{E}}_w$ for different values of $\lambda _{abs} $, which can be compared to the $y$-component of $\mathcal{E}_w = 1/d = 33.3$~cm$^{-1}$ for $\lambda_{abs} \gg p$.
    }
  \label{fig:Eweff}
 \end{figure}

 Fig.~\ref{fig:Rlambda} shows  $R_y(y)$, the ratio of the $y$-component of  $\overrightarrow {\mathcal{E} _w}$ for different values of $\lambda _{abs}$ to the one for $\lambda _{abs} = 1~\mathrm{m}$, i.~e. $\lambda _{abs} \gg p$.
 It can be seen that for $\lambda _{abs} = 2$~mm, which is close to the value for the laser light used for the measurements, $R_y = 1$ within $\pm~1$~\%, and the deviation from 1 is small compared to other uncertainties of the method.
 Fig.~\ref{fig:Ex-weight} shows the $x$-component of $\overrightarrow {\mathcal{E} _w}$ for different values of $\lambda _{abs}$.
 For small $y$, values as high as $\approx 5$~cm$^{-1}$ are found for $\lambda _{abs} = 2$~mm, which corresponds to $\approx 15$~\% of the $y$-component of $\mathcal{E} _w$.
 Its effect on the analysis depends on the electric field distribution in the region of the readout plane, which is not known.
 No further study has been made on this topic.

 \section{Appendix B: Influence of  electronics and  charge carrier lifetimes on the velocity profiles}
  \label{sect:AppendixB}

 In this appendix simulations are performed with the aim to investigate the influence of the finite bandwidth of the readout electronics and the finite lifetimes of electrons and holes on the determination of the electric field from the velocity profiles.

 In order to use electric fields in the simulations, which are similar to the ones shown in Fig.~\ref{fig:Efit-irr-rev-1E16}, a pad sensor with a doping density of $2 \times 10^{13}$~cm$^{-3}$ has been simulated.
 The corresponding depletion voltage is $\approx 1400$~V for a $300~\upmu$m thick sensor.
 In the non-depleted region of the sensor the electric field is assumed to be 1~kV/cm because of the resistivity of the radiation-damaged silicon and the dark current.
 No increase of the field towards $y = d$, the \emph{double junction}, is implemented, as for the irradiated detectors studied  the effect is minor.
 In the following  results for two voltages, 200~V and 1000~V, are shown, for which the maximum fields at $y=0$ are 35~kV/cm and 80~kV/cm, respectively.
 The depletions depths for these two voltages are $115~\upmu$m and $275~\upmu$m.
 It has been checked that the conclusions from these simulation are the same if these quite arbitrary assumptions are changed.

\begin{figure}[!ht]
   \centering
   \begin{subfigure}[a]{0.5\textwidth}
    \includegraphics[width=\textwidth]{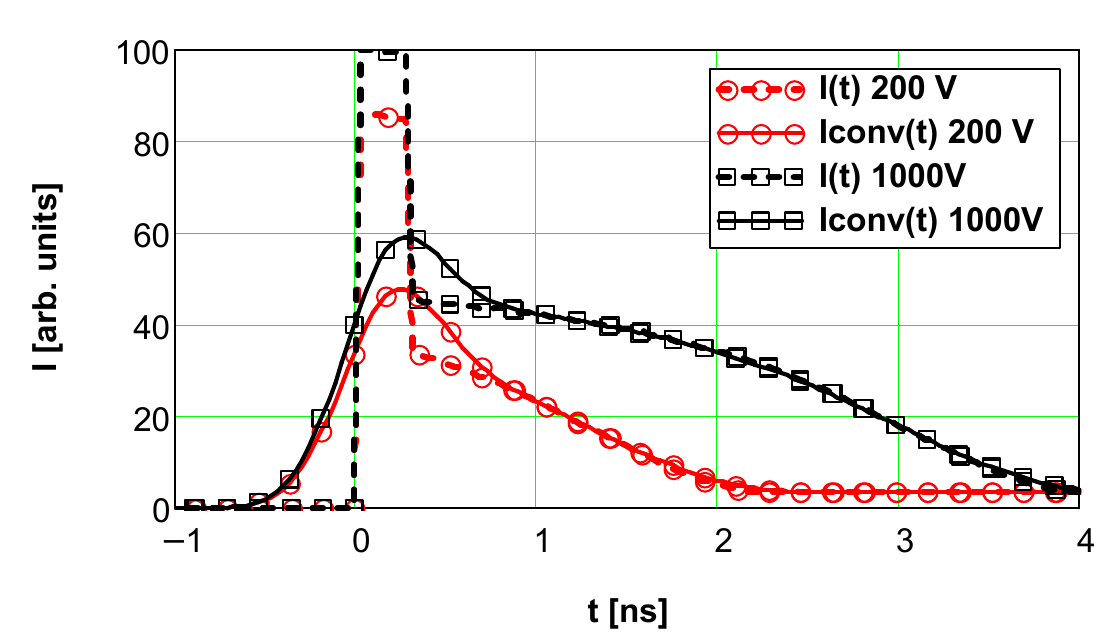}
    \caption{ }
    \label{fig:Trans25mum}
   \end{subfigure}%
   \begin{subfigure}[a]{0.5\textwidth}
    \includegraphics[width=\textwidth]{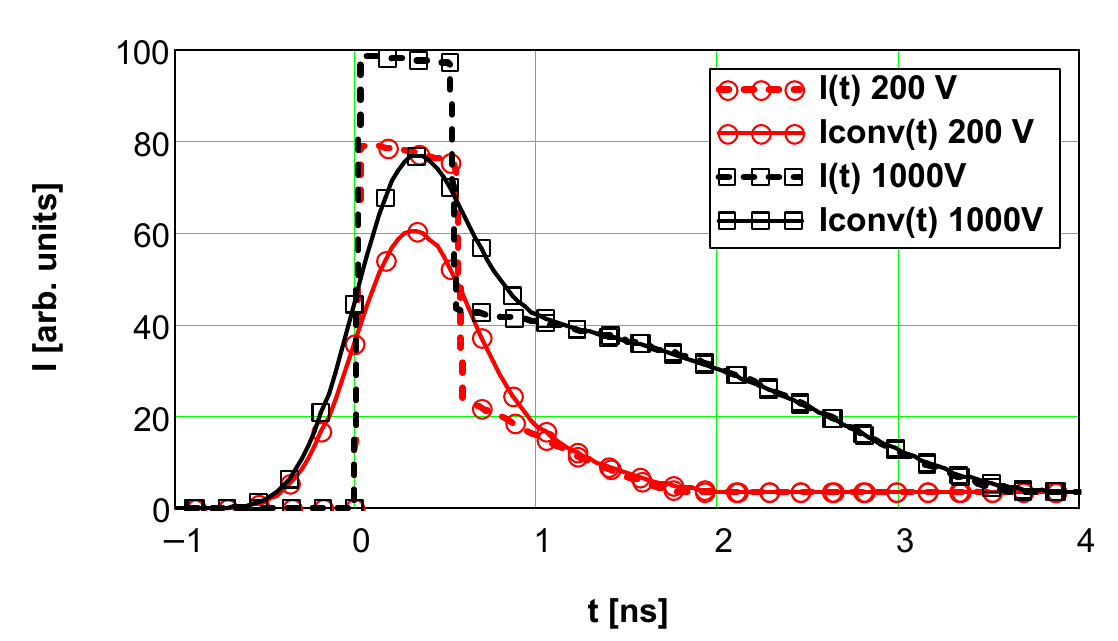}
   \caption{ }
    \label{fig:Trans50mum}
   \end{subfigure}%
 \newline
   \begin{subfigure}[a]{0.5\textwidth}
    \includegraphics[width=\textwidth]{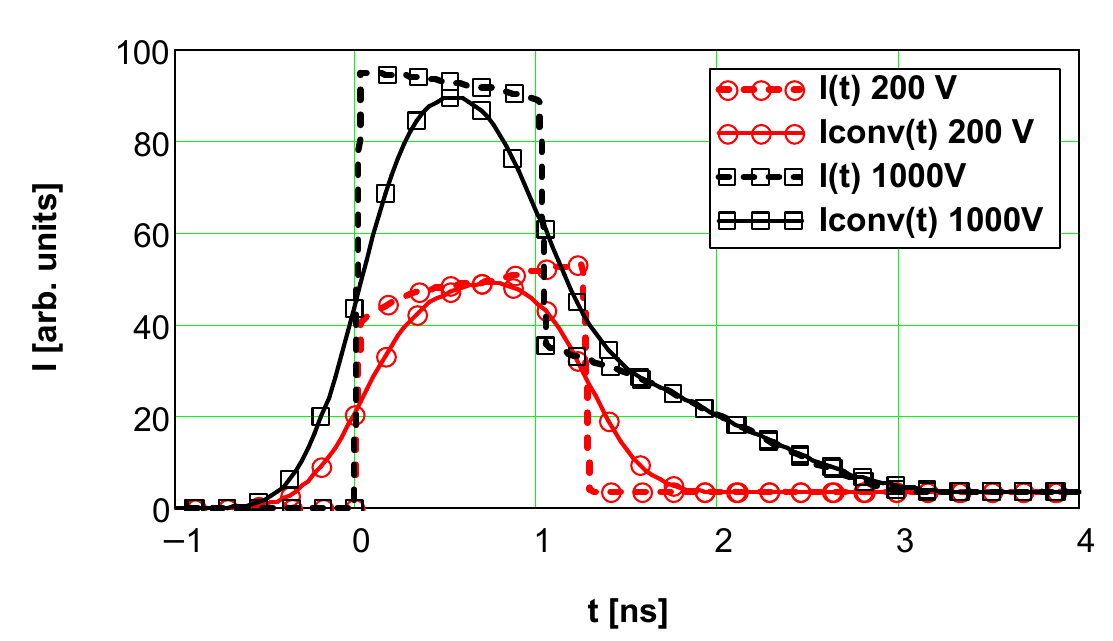}
    \caption{ }
    \label{fig:Trans100mum}
   \end{subfigure}%
   \begin{subfigure}[a]{0.5\textwidth}
    \includegraphics[width=\textwidth]{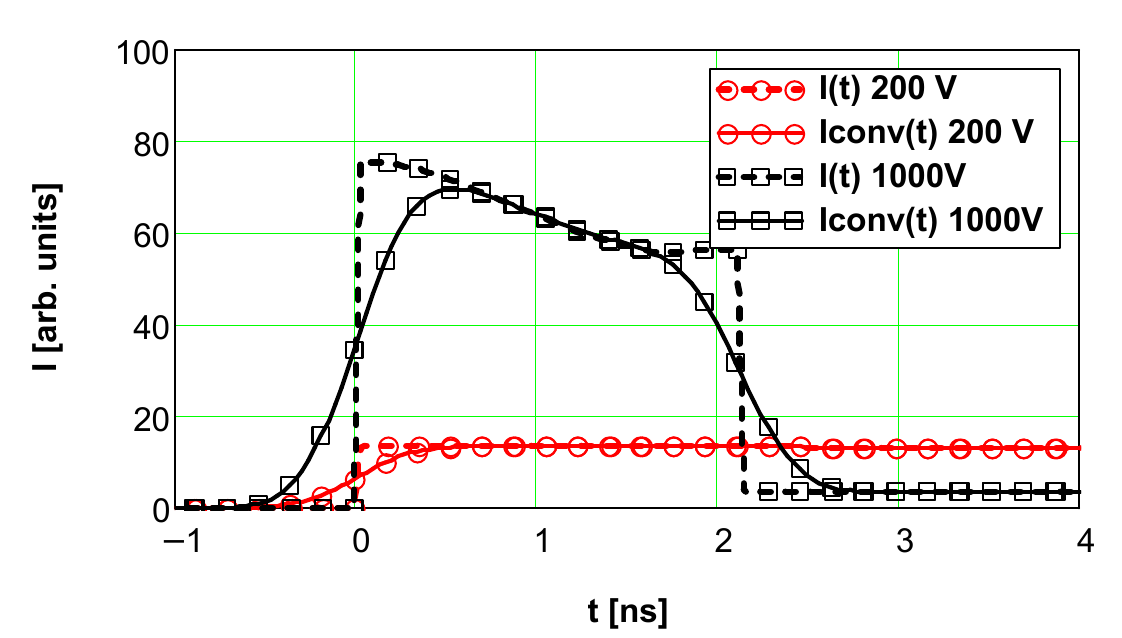}
   \caption{ }
    \label{fig:Trans200mum}
   \end{subfigure}
   \caption{
    Current transients for the simulations discussed in the text for electron-hole~pairs generated uniformly in $x$ at $t=0$ and different $y$~values.
    The transients before (dashed lines) and after (solid lines) the convolution with a Gaussian with $\sigma _t = 250$~ps for the voltages $U= 200$~V and $U=1000$~V are shown for
      (a) $y = 25~ \upmu $m,
      (b) $y = 50~ \upmu $m,
      (c) $y = 100~ \upmu $m, and
      (d) $y = 200~ \upmu $m.
    }
  \label{fig:Trans}
 \end{figure}

 For the effective weighting field $\hat e_y/d$ is assumed.
 For the drift of the charge carriers the mobility parametrisation of Ref.~\cite{Scharf:2015} is used and effects of charge diffusion are neglected.
 With the help of Eq.~\ref{equ:Igeom} the time dependence of the induced current $I(t;y_0)$ for a uniform chain of electron-hole~pairs produced at $ y_0$ is calculated.
 Examples for the generated current transients are shown in Fig.~\ref{fig:Trans} as dashed lines.

 To investigate the influence of the electronic response function on the initial rise of the measured transient, the simulated current transient, $I(t)$, is convolved with a Gaussian response function with an rms time spread $\sigma _t$.
 Both functions are evaluated at 512 time steps in the interval between $-3$~ns to $+15$~ns, resulting in a time step of $\approx 35$~ps.
 The electron-hole~pairs are generated at $t = 0$, and the Gaussian response function is also centred at $t = 0$.
 The method of the Fast-Fourier-Transform (FTT) is used for the convolution.
 For the estimation of $\sigma _t$ the convolved current simulated for the non-irradiated sensor at $y = 50~\upmu$m is compared to the measured current shown in Fig.~\ref{fig:Trans_nonirr}.
 The value of $\sigma _t$ is varied until the rise time of the convolved current agrees with the measured value.
 The value found is $\sigma _t \approx 250$~ps, which corresponds to a bandwidth $BW = 0.0935 \cdot \sqrt{2}/\sigma _t \approx 530$~MHz and a rise time $t_r = 0.3394/BW \approx 640$~ps.

 Fig.~\ref{fig:Trans} shows several examples of current transients before and after convolution.
 It is noted that for small $y$~values the electrons which drift to $y = 0$ produce a narrow flat region, which is strongly suppressed by the convolution.
 As a result the maximum of the convolved transient is much lower than the original transient.
 For larger values of $y$ the width of the electron signal becomes wider than the full width, $\Gamma _t$, of the Gaussian response function, and the maxima of the transients before and after convolution become similar.
 We thus expect that the determination of the electric field from the velocity profiles will only be reliable for $y > v_e \cdot \Gamma _t$, where $v_e$ is the electron velocity close to $y = 0$.
 A similar effect from holes is expected at $y = d$.
 However, for the neutron-irradiated sensors studied in this paper, the electric field is much lower in this region than at $y = 0$, and the effects are much smaller.
 In principle this effect can be taken into account in the analysis, which, however, has not been done.

 Two methods of determining the velocity profiles from the convolved transients are investigated for the simulated data:
 \begin{enumerate}
   \item The slope method, where it is found that taking the maximum slope or the slope at $t = 0$ makes hardly a difference.
   \item The time integral method, for which the transient is integrated up to the time $t_{int}$.
 \end{enumerate}

 \begin{figure}[!ht]
   \centering
   \begin{subfigure}[a]{0.5\textwidth}
    \includegraphics[width=\textwidth]{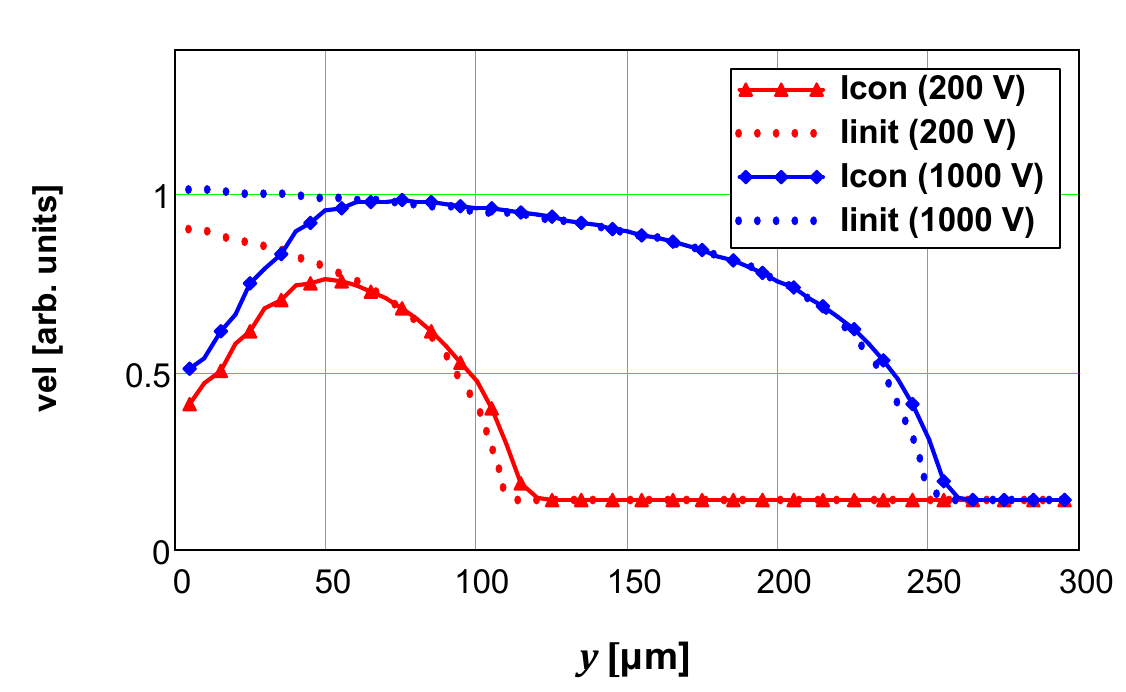}
    \caption{ }
    \label{fig:VelSlope}
   \end{subfigure}%
   \begin{subfigure}[a]{0.5\textwidth}
    \includegraphics[width=\textwidth]{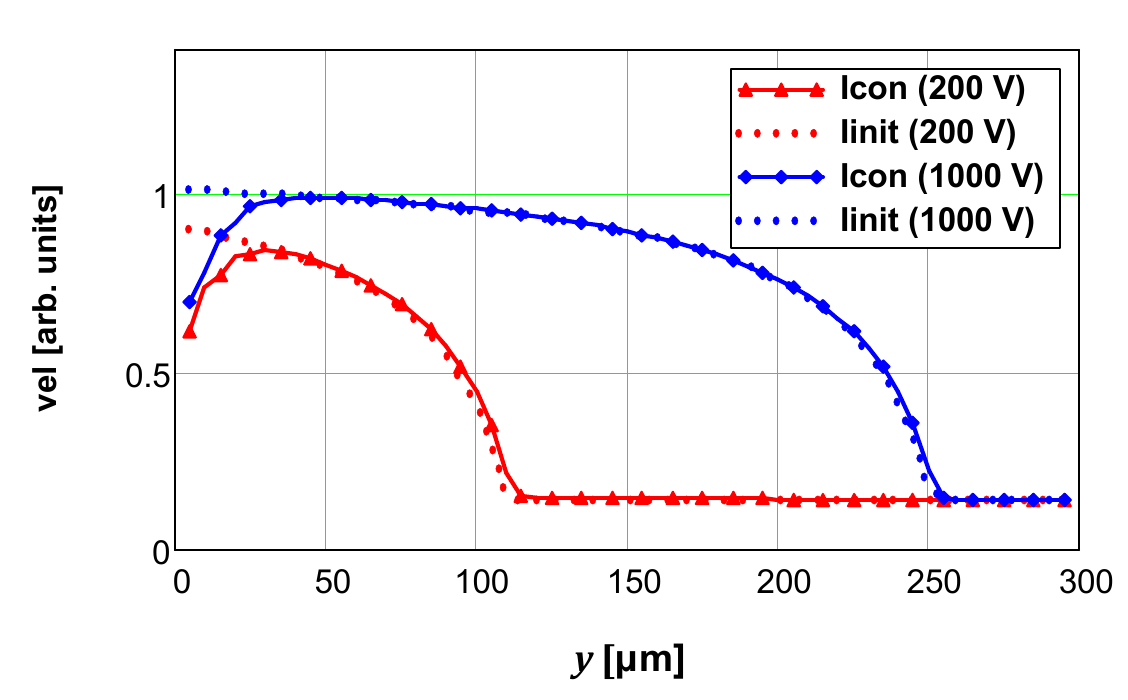}
   \caption{ }
    \label{fig:VelInt}
   \end{subfigure}
   \caption{Comparison of  velocity profiles before (dotted lines) and after (solid lines) the convolution of the simulated current transients for $U = 200$~V and $U = 1000$~V for
     (a) the slope method, and
     (b) the charge integration method for $t_{int} = - 250$~ps.
    }
  \label{fig:VelSim}
 \end{figure}

 Fig.~\ref{fig:VelSim} compares the results of the two methods.
 For the slope methods significant deviations between the simulated velocity profile and the velocity profile derived after the convolution appear below $y = 50~\upmu$m.
 Differences are also observed at the transition between the depleted and the non-depleted region ($y = 120~\upmu$m for 200~V, and $y = 260~\upmu$m for 1000~V).
 For the charge integration method, the integration time $t_{int}$ can be optimised.
 It is found that for $t_{int} \gtrsim 250$~ps the results are identical to method 1. However, for $t_{int} = - 250$~ps, i.~e. the integration of the convolved transient starts $\sigma _t$ before the pulse without convolution, the agreement can be extended down to $\approx 30~\upmu$m and the deviations when approaching the non-depleted region disappears.
 However such a short integration time results in significant experimental errors, and thus cannot be used in the analysis.

 For the study of the effects of the finite lifetimes of the free charge carriers, $\tau _i $,  the numbers of the drifting charges are reduced following an exponential in time, and the resulting velocity profiles are compared to the velocity profiles for  $\tau _i $~values large compared to $t_{trans}$, the transit time of the charge carriers.
 For both methods, where $t_{int} = + 250$~ps is used for method (2), the results are similar and, assuming $\tau = \tau_e = \tau_h $, it is found from the simulations that the relative reduction of $vel$ is approximately $\sigma _t/ \tau = 0.25~\mathrm{ns}/\tau$, with 
 $\sigma _t$ the rms width of the Gaussian transfer function introduced before.
 As the $\tau _i$~values of silicon irradiated to fluences of $\Phi _{eq} = 10^{16}$~cm$^{-2}$ are only a few ns, the effect is significant.
 However, in particular if the $\tau _i$~values are independent of position, the reduction of $vel$ is similar in the entire sensor and partially compensated by the requirement $\int _0 ^d E(y)~\mathrm{d}y = U$.

 To summarise this Appendix:
 The simulations show that both, the finite bandwidth of the readout and free charge-carrier lifetimes of a few nanoseconds and less, significantly influence the determination of the electric field from velocity profiles.
 Whereas the  effect of the finite bandwidth is limited to the edges of the sensor, short lifetimes affect the electric field determination in the entire sensor.
 To avoid these uncertainties the entire transients and not only its initial rise should be used in the analysis.
 However, this requires the precise understanding of the electronic transfer function, which could be possibly achieved using the method described in Refs.~\cite{Scharf:2014, Scharf1:2015}.


\end{document}